\documentclass[conference]{IEEEtran}
\usepackage{epsfig,endnotes}

\pdfoutput=1
\usepackage{silence}
\usepackage[hyphens]{url}
\usepackage[breaklinks,colorlinks]{hyperref}
\usepackage[usenames,dvipsnames]{xcolor}
\hypersetup{citecolor=blue,linkcolor=blue,urlcolor=black}
\usepackage{amsmath,amsopn,amssymb}
\usepackage{subcaption}
\usepackage{endnotes,microtype,xspace,graphicx,fancyvrb,multirow}
\usepackage{booktabs}
\usepackage{array,underscore,relsize}
\usepackage[T1]{fontenc}
\usepackage{times}
\usepackage{fancyhdr,lastpage}

\usepackage{enumitem}
\usepackage[labelfont=bf,font=small,skip=5pt]{caption}
\usepackage{tabularx}
\usepackage{fp}
\usepackage{siunitx}
\usepackage{comment}

\usepackage{balance}

\usepackage{algpseudocode}
\usepackage[ruled,vlined,linesnumbered,nofillcomment]{algorithm2e}
\SetAlFnt{\footnotesize}

\SetCommentSty{mycommfont}

\usepackage[square,comma,numbers,sort&compress]{natbib}
\newcommand{\sys}{\mbox{\textsc{SynFuzz}}\xspace}

\newcommand{\cc}[1]{\mbox{\smaller[0.5]\texttt{#1}}}

\fvset{fontsize=\scriptsize,xleftmargin=8pt,numbers=left,numbersep=5pt}

\input{code/fmt}

\def\Snospace~{\S{}}

\newif\ifdraft\drafttrue
\newif\ifnotes\notestrue
\ifdraft\else\notesfalse\fi

\input{glyphtounicode}
\newcolumntype{R}[1]{>{\raggedleft\let\newline\\\arraybackslash\hspace{0pt}}p{#1}}

\newcommand{\squishlist}{
\begin{itemize}[noitemsep,nolistsep]
  \setlength{\itemsep}{-0pt}
}
\newcommand{\squishend}{
  \end{itemize}
}

\usepackage{xstring}
\newcommand{\PP}[1]{
\vspace{2px}
\noindent{\bf \IfEndWith{#1}{.}{#1}{#1.}}
}

\newcommand{\etal}{\textit{et al}.\xspace}
\newcommand{\ie}{\textit{i}.\textit{e}.}
\newcommand{\eg}{\textit{e}.\textit{g}.}
\newcommand{\aka}{\textit{a}.\textit{k}.\textit{a}.}

\begin{document}

\title{\sys: Efficient Concolic Execution via Branch Condition Synthesis}

\ifdefined\DRAFT
 \pagestyle{fancyplain}
 \lhead{Rev.~\therev}
 \rhead{\thedate}
 \cfoot{\thepage\ of \pageref{LastPage}}
\fi


\author{
\IEEEauthorblockN{
    Wookhyun Han\IEEEauthorrefmark{1},
    Md Lutfor Rahman\IEEEauthorrefmark{2},
    Yuxuan Chen\IEEEauthorrefmark{3},
    Chengyu Song\IEEEauthorrefmark{2},
    Byoungyoung Lee\IEEEauthorrefmark{4},
    \\
    Insik Shin\IEEEauthorrefmark{1}
}
\\
\IEEEauthorblockA{
    \IEEEauthorrefmark{1}KAIST,
    \IEEEauthorrefmark{2}UC Riverside,
    \IEEEauthorrefmark{3}Purdue,
    \IEEEauthorrefmark{4}Seoul National University
}
}

\date{}
\maketitle

\begin{abstract}
Concolic execution is a powerful program analysis technique for
exploring execution paths in a systematic manner.
Compare to random-mutation-based fuzzing,
concolic execution is especially good at exploring paths that are
guarded by complex and tight branch predicates (\eg, \cc{(a*b) == 0xdeadbeef}).
The drawback, however, is that concolic execution engines are much
slower than native execution.
One major source of the slowness is that concolic execution engines
have to the interpret instructions to maintain the symbolic expression
of program variables.
In this work, we propose \sys, a novel approach to perform scalable
concolic execution.
\sys achieves this goal by replacing interpretation with dynamic taint
analysis and program synthesis.
In particular, to flip a conditional branch, \sys first uses
operation-aware taint analysis to record a partial expression (\ie,
a sketch) of its branch predicate.
Then it uses oracle-guided program synthesis to reconstruct the
symbolic expression based on input-output pairs.
The last step is the same as traditional concolic execution---\sys
consults a SMT solver to generate an input that can flip the target
branch.
By doing so, \sys can achieve an execution speed that is close to
fuzzing while retain concolic execution's capability of flipping
complex branch predicates.
We have implemented a prototype of \sys and evaluated it with three sets
of programs: real-world applications, the LAVA-M benchmark, and
the Google Fuzzer Test Suite (FTS).
The evaluation results showed that \sys was much more scalable than
traditional concolic execution engines,
was able to find more bugs in LAVA-M than most state-of-the-art concolic execution engine (QSYM),
and achieved better code coverage on real-world applications and FTS.
\end{abstract}

\section{Introduction}
\label{s:intro}

\begin{table*}[t]
  \centering
  \caption{Bugs found on the LAVA-M data set by different testing
  techniques. \sys outperforms most other techniques and found many
  bugs that are not listed by LAVA authors. For other techniques,
  we used the number reported in the corresponding papers.}
\scriptsize
\begin{tabular}{lrrrrrrrrrrrrrrr}
\toprule
  Program & Listed bugs & \textbf{\sys} & \textsc{Redqueen} & \textsc{Neuzz}
  & Angora & QSYM & Eclipser & ProFuzzer & T-Fuzz & Steelix & VUzzer & AFL & Fuzzer & SES \\
\midrule
uniq   &   28 &   29 &   29 &   29 &   29 &   28 &   29 & 28 & 29 &   7 & 27 & 9 & 7 &  0 \\
base64 &   44 &   48 &   48 &   48 &   48 &   44 &   46 & 24 & 47 &  43 & 17 & 0 & 7 &  9 \\
md5sum &   57 &   61 &   61 &   60 &   57 &   57 &   55 & 10 & 55 &  28 &  0 & 0 & 2 &  0 \\
who    & 2136 & 2218 & 2462 & 1582 & 1541 & 1238 & 1135 & 41 & 70 & 194 & 50 & 1 & 0 & 18 \\
\bottomrule
\end{tabular}

  \label{tbl:lava-m}
\end{table*}


Concolic execution is a very powerful program analysis technique
\cite{godefroid2008automated,chipounov2011s2e,cadar2008exe,yun:qsym,
sen2005cute,cadar2013symbolic,bounimova2013billions,
godefroid2005dart,cadar2008klee,cha2012unleashing}.
When used in software testing or bug finding,
we mark untrusted inputs as symbolic values and utilize the concolic
execution engine to perform path exploration.
During the execution, the engine maintains (i) symbolic values, which
represents program variables (memory and registers content) as
symbolic expressions, and (ii) a set of symbolic constraints over the
symbolic values, which are imposed by conditional branches that have
been taken along the execution path~\cite{cadar2013symbolic}.
Whenever the engine encounters a new conditional branch where the
branch predicate is symbolic,
it constructs a boolean formula for each branch target condition
(\ie, $predicate = \mathtt{true}$ for the true branch and
$predicate = \mathtt{false}$ for the false branch).
Then it consults a satisfiability modulo theories (SMT) solver to
check whether the branch condition is satisfiable.
If the branch condition is satisfiable,
the engine forks a new state (so as to explore both branches),
adds this new path constraint to the constraint set and continues
the path exploration as if the target branch has been taken.
To generate a concrete input that can follow the same path,
the engine can ask the SMT solver to provide a feasible assignment to
all the symbolic input bytes that satisfy the collected path constraints.
Thanks to the recent advances in SMT solvers, compare to random
mutation based fuzzing, concolic execution is known to be much more
efficient at exploring paths that are guarded by complex and tight
branch predicates (\eg, \cc{(a*b) == 0xdeadbeef}).

However, a critical limitation of concolic execution is its scalability
\cite{yun:qsym,cadar2013symbolic,bounimova2013billions}.
More specifically, a concolic execution engine usually takes a very
long time to explore ``deep'' execution paths.
The slowness comes from two sources.
The first source is the SMT solver.
The SMT problem is a generalized version of the Boolean Satisfiability
problem (SAT) which is proven to be NP-complete.
As the accumulated path constraints grow, solving a new branch
condition using the SMT solver becomes slower
\cite{cadar2013symbolic,liu2014comparative,yun:qsym}.
The second source of overhead is symbolic interpretation of instructions.
In order to maintain symbolic expressions,
most concolic execution engines interpret the target program
instruction-by-instruction,
which is much slower than native execution.
For example, Yun~\etal~\cite{yun:qsym} reported that
KLEE~\cite{cadar2008klee} is around 3,000 times slower than native execution and
angr~\cite{shoshitaishvili2016sok} is more than 321,000 times slower.
As a result, even with a targeted path to explore~\cite{ma2011directed,
godefroid2005dart}
(\eg, forcing the execution to follow a given execution trace up to a certain point),
it could still take a concolic execution engine hours of time to finish
interpreting the execution trace.
Due to the scalability issue stemming from these two sources,
concolic execution is not as effective as greybox fuzzing in
finding bugs in real-world programs.
For example, all DARPA Cyber Grand Challenge finalist teams used
both fuzzing and concolic execution to find bugs but most bugs were
found by fuzzers~\cite{stephens2016driller}.


In this work, we aim to improve the scalability of concolic execution
by replacing symbolic interpretation with
\emph{dynamic taint analysis}~\cite{newsome2005dynamic}
and \emph{program synthesis}~\cite{gulwani2017program}.
More formally, let $T = \{b_1,b_2,\dots,b_n\}$ be the branch trace of
executing a program $P$ with a vector of bytes $\mathbf{x}$ as input.
For each branch $b_i$ that is directly controlled by input,
we denote $f_i(\mathbf{x}_i)$ as its predicate
(\ie $f_i$ is a boolean function evaluated to either $\mathtt{true}$
or $\mathtt{false}$, and $\mathbf{x}_i$ 
are the relevant input bytes that directly control the branch $b_i$).
By performing a byte-level dynamic taint analysis,
we can know both $\mathbf{x}_i$ and the corresponding evaluation
result of $f_i(\mathbf{x}_i)$.
However, to obtain a new input ${\mathbf{x}_i}'$ that can ``flip''
the branch $b_i$ (\ie, $f_i({\mathbf{x}_i}') \neq f_i(\mathbf{x}_i)$)
with a SMT solver,
we need to know the ``body'' of $f_i$.

Traditional concolic execution engine solves this problem through
construction---by interpreting executed instructions symbolicly,
it keeps track of how $f_i$ is formulated.
In this work, we use program synthesis to derive the body of $f_i$.
Specifically, given a black-box function $f_i$, how to reconstruct a
function $f_i^{syn}$ that generates the same output as $f_i$ when
provided the same input is a typical oracle-guided program synthesis
problem~\cite{jha2010oracle}.
Therefore we (i) use dynamic taint analysis to figure out relevant bytes
($\mathbf{x}_i$);
(ii) mutate these bytes to collect new input-output pairs;
and (iii) leverage program synthesis to derive $f_i^{syn}$.
Once we synthesized a candidate $f_i^{syn}$ using existing input-output pairs,
we can ask a SMT solver to provide the corresponding ${\mathbf{x}_i}'$ that
would allow us to explore the other branch.
Then we test this new input $\mathbf{x}'$ with the original program
(\ie, the oracle):
if the input indeed flips the evaluation of $b_i$,
this means our synthesis result is correct;
otherwise we add this new input-output pair into the set and continue
searching for the correct $f_i^{syn}$.
By doing so, we can enjoy an execution speed that is close to fuzzing while
retain concolic execution's capability of flipping complex branch predicates.
Note that similar to Angora~\cite{angora},
we only need to perform the expensive dynamic taint once to capture $\mathbf{x}_i$,
generating additional input-output pairs.
Then testing synthesis results can be done without taint tracking.


The challenge of the aforementioned approach, however, is that the
search space for the synthesis grows factorially over the size of $f_i$.
This means a naive, enumeration-based synthesis algorithm
(see~\autoref{s:background} for details) could become very slow
and overshadows the performance benefit of native execution.
To overcome this challenge, we drastically reduce the search space by
logging extra computational history of execution (\ie, a sketch)
during the dynamic taint analysis.
Even though we logged extra information,
it is still scalable to perform dynamic taint analysis
for real-world applications.
We observe that, when synthesizing a new line of code,
the synthesizer needs to iterate through
(i) possible operations and (ii) possible operands,
which could be the input bytes, concrete values, or results of
previous operations/lines.
By limiting the possible choices of operations and operands,
we can make the search much more efficient.

Based on this observation, \sys extends taint labels to include both
operations and operands.
%
In particular, each taint table is a tuple $(op, operand1, operand2, size)$,
where $op$ is the operation that caused the two labels to merge,
$operand1$ and $operand2$ are the corresponding labels that are merged,
and $size$ is the size of the operands.
An operand could be the offset of input byte (leafs) or
the index of another taint table entry.
This essentially allows us to maintain an AST-like computation history of a
taint label, yet with a major difference:
we do not record the value of any concrete (untainted) variable.
Instead, we use a simple placeholder (label $0$) to indicate the
corresponding operand is concrete.
We made this design choice because
(i) the potentially enormous number of unique concrete values may quickly
exhaust the taint labels; 
and (ii) in most cases, actual values of these concrete variables can be
easily recovered during synthesis.
In other words, we use dynamic taint analysis to record as much
information as possible and rely on program synthesis to recover
the lost information.

While we were implementing this idea, we also solved additional challenges.
%
The first challenge is how to identify $b_i$ so we can record the
corresponding input-output pairs.
A simple idea, as used by the AFL fuzzer~\cite{afl}, is to use the
address of $b_i$ as its ID.
However, since the same branch instruction may be executed multiple
times under different \emph{context},
the symbolic expression of its predicate would also be different.
As a result, this method could prevent the synthesizer from
constructing the predicate because we are mixing input-output pairs
for different $f_i$ together
(see~\autoref{s:discuss} for more details).
Ideally, we should identify $b_i$ using its position in the execution
trace $T$;
but as the trace $T$ could be very large, recording and parsing
the full execution trace could be very expensive.
In our current prototype, we used the call stack and the incoming branch
to represent the context as they provide a reasonable trade-off
between accuracy and scalability.
%
%

The second challenge is how to handle nested branches,
\ie, when trying to flip a target branch $b_t$,
we want to preserve the outcome of all precedent branches
($b_i, i \in [0, t)$) so the trace can still reach $b_t$.
Similar to traditional concolic execution~\cite{godefroid2005dart,
godefroid2008automated,cadar2008exe},
we solve this challenge by maintaining/remembering the synthesized
symbolic predicate $f_i$ for each $b_i$ and force them to follow
the same branch direction as in the trace.
To achieve this goal, we log tainted branch conditions in a
chronological order; then we process these branches in the same order.
To process a branch, we first check if the opposite branch is satisfiable;
if so, we create a new concrete input.
Then we add a constraint to the solver to force this branch to follow
the original direction.
By doing so, when processing a target branch $b_t$,
we also enforce all precedent branches affected by
the input bytes $\mathbf{x}_t$ to follow the same paths.

Compared with taint-analysis-guided fuzzing,
\sys has two main advantages.
First, its ability to reconstruct complex branch predicate allows it to
handle more versatile predicates other than magic number testing
\cite{vuzzer,li2017steelix,lemieux2017fairfuzz,you2019profuzz,wang2010taintscope},
simple transformations~\cite{aschermann2019redqueen},
and monotonic functions~\cite{angora,szekeres:thesis,she2018neuzz,choi2019grey}.
Second, its ability to reconstruct and maintain symbolic predicates
allows it to consider multiple branches and avoid accidentally flipping
earlier branches in the trace.


We have implemented a hybrid fuzzing system with \sys and AFL.
We evaluated \sys with real-world applications,
the LAVA-M~\cite{dolan2016lava} benchmark, and
the Google Fuzz Test Suite (FTS)~\cite{FTS}.
On real-world applications and the FTS, we achieved better coverage
than AFLFast~\cite{bohme2016coverage} and QSYM~\cite{yun:qsym}.
On LAVA-M data set (\autoref{tbl:lava-m}),
\sys outperformed most other state-of-art fuzzers and
symbolic/concolic execution engines\footnote{Based on reported numbers}.
On FTS, we were able to find bugs faster than AFLFast in most cases.

In summary, this paper makes the following contributions:

\begin{itemize}
	\item \textbf{New test generation technique.} We proposed a novel
	approach to perform formula-solving-based test generation by
	combining dynamic taint analysis and program synthesis.

	\item \textbf{Efficient fine-grained taint tracking.}
	To achieve efficient synthesis,
	we design and implement an efficient byte-level dynamic taint
	tracking system to record detailed information about how input
	bytes are affecting the branch condition.

	\item \textbf{Hybrid fuzzer.} As a prototype,
	we implement a hybrid fuzzing system \sys and applied it to two
	fuzzing benchmark suites with real-world applications.
	The evaluation results showed that \sys outperformed all other
	state-of-the-art testing techniques.

\end{itemize}

\section{Background and Motivations}
\label{s:background}

In this section, we provide a quick overview of fuzzing, symbolic
execution, and program synthesis.
Then we highlight the motivation for this work.

\subsection{Background: Fuzzing}

Fuzzing is a dynamic testing technique that tries to trigger software
bugs with randomly generated inputs.
Based on how the inputs are generated,
fuzzers can be categorized into generational fuzzers
and mutational fuzzers.
Generational fuzzers can be grammar guided~\cite{godefroid2008grammar,
peachfuzzer,aitel2002introduction,jsfunfuzz}
or learning based~\cite{wang2017skyfire,rajpal2017not,godefroid2017learn}.
Mutational fuzzers create new inputs using random mutation
and crossover~\cite{miller1990empirical,afl,honggfuzz,wang2010taintscope,vuzzer}.
Based on how much runtime information is gathered and used,
fuzzers can be categorized into blackbox, greybox, and whitebox fuzzers.
Blackbox fuzzers~\cite{miller1990empirical,peachfuzzer,aitel2002introduction}
do not consider any runtime information to assist generating new inputs.
Whitebox fuzzers~\cite{godefroid2008automated,godefroid2005dart,
cadar2008exe,yun:qsym,vuzzer,angora}
use heavyweight program analysis techniques
(\eg, dynamic taint analysis and symbolic execution)
to collect detailed information to assist input generation.
Greybox fuzzers~\cite{afl,honggfuzz,lemieux2018perffuzz,petsios2017slowfuzz,
aschermann2019redqueen,you2019profuzz,choi2019grey}
use lightweight instrumentation to collect a small amount of
runtime information to guide input generation,
such as coverage information~\cite{afl,honggfuzz,shen2019neuro}
and performance information~\cite{lemieux2018perffuzz,petsios2017slowfuzz}.
Since greybox and whitebox fuzzers have a feedback loop to guide
new inputs generation, they are also called evolutionary fuzzers.
Despite their simplicity, fuzzers are quite effective at finding
bugs~\cite{ossfuzz}.

\subsection{Background: Dynamic Symbolic Execution}

Symbolic execution treat program inputs as symbolic values instead of
concrete values.
Program variables (including memory and register content)
can then be represented as symbolic expressions,
\ie, functions of symbolic inputs.
A symbolic execution engine maintains (i) a symbolic state $\sigma$,
which maps program variables to their symbolic expressions, and
(ii) a set of path constraints $PC$, which is a quantifier-free
first-order formula over symbolic expressions~\cite{cadar2013symbolic}.
To generate a concrete input that would allow the program to follow
the same execution trace, the symbolic execution engine uses $PC$ to
query an SMT solver for satisfiability and feasible assignment to
symbolic values (\ie, input).
The advantage of symbolic execution over random mutation/generation
is the ability to handle complex branch conditions more efficiently.
The drawback, however, is the lack of scalability.
As mentioned in the introduction,
this is mainly caused by two reasons: solver and symbolic interpretation
\cite{yun:qsym,cadar2013symbolic}.

\subsection{Background: Program Synthesis}

Program Synthesis is the task of automatically finding programs that
satisfy user-specified constraints~\cite{gulwani2017program}.
Program synthesizers typically perform some form of search over the
space of programs to generate the program that satisfies the constraints.
The constraints can be expressed as input-output examples, demonstrations,
natural language, partial programs, and mathematical logics.
The search can be enumerative, deductively guided, or constraint-guided.
Since the search space is extremely huge, many heuristics have been
proposed to prune the search space~\cite{udupa2013transit}
or improve the likelihood of finding the correct program
\cite{le2012genprog,schkufza2013stochastic},
including utilizing neural network~\cite{lee2018accelerating,
balog2016deepcoder,feng2018program,si2018learning}.

In this work, we use oracle-guided, sketch-based, and solver-assisted
search for branch condition synthesis.
In particular, given (i) a vector of input bytes $\mathbf{x}$,
(ii) a set of concrete variables $C$, and
(iii) a set of possible operations $O$,
we enumerate functions $f_0,f_1,\dotsc,f_n$ as unique combinations of them.
Each function $f_i$ has $l$ lines where each line $l_j$ has
a single operation $o_j \in O$, a single output $out_j$,
and one or more operands $in_j$ depending on the corresponding operation $o_j$.
Operands can be input byte(s),
concrete variable(s) $c \in C$,
or outputs from previous lines $out_k, k<j$.
Input bytes, concrete variables, and outputs can be used in multiple lines.
Once we have constructed a function $f_i$,
we add observed input-output pairs as constraints over $f_i$:\\
\[
assert(f_i(\mathbf{x}) = y),\quad y \in \{\mathtt{true,false}\}
\]
\\
and ask the solver to provide a feasible assignment of concrete
variables $c \in C$ that would satisfy the constraints.
Note that the output has only one value because this work is only interested
in conditional branches that are always evaluated to $\mathtt{true}$ or
$\mathtt{false}$.
If the solver is able to find an assignment,
we have a candidate function.
We then ask the solver to provide a feasible assignment of input bytes
$\mathbf{x}'$ that would yield the opposite output ($\neg y$).
Finally, we verify the correctness of $f_i$ by executing the original
program $P$ with the new input.
If the new input indeed flips the branch, we believe $f_i$ is the
correct function until a new counterexample is discovered;
otherwise we add the new input-output pair to the observed set and
continue the search.

\subsection{Motivation}

For coverage-driven testing techniques, the ultimate goal is to visit
both the true and false branches of every conditional branch,
\ie, 100\% branch coverage.
To achieve this goal, we need the ability to flip conditional branches.
For a conditional branch $b_i$ that can be directly affected by inputs,
this goal can be achieved in two general ways.
First, we can treat the corresponding $f_i$ as a blackbox and try to
flip its output without understanding how $f_i$ maps inputs to the output.
Methods belong to this category include
random guessing (\ie, fuzzing)~\cite{miller1990empirical,afl,honggfuzz},
simulated annealing~\cite{szekeres:thesis},
binary search~\cite{choi2019grey}
gradient descent~\cite{angora,she2018neuzz}, etc.

The second way is to ``derive'' $f_i$ (\ie, as a whitebox)
and then solve it mathematically.
The most popular method in this category is dynamic symbolic execution,
where we derived $f_i$ through symbolic interpretation and change its output using
an SMT solver.
Another method researchers have tried is to simulate $f_i$ using a deep
neural network and flip the output using gradient descent~\cite{shen2019neuro}.
Whitebox approaches are more powerful than blackbox approaches but
the process to derive $f_i$ could be very slow, making them less
efficient than blackbox approaches.

To speedup the deriving process, we make the following key
observations:
\begin{enumerate}
	\item For most branches, $f_i$ is not complex in terms of
	(i) the number of input bytes it depends on
	(\ie, the size of the input $\mathbf{x}_i$) and
	(ii) the complexity of the computation
	(\ie, the size of the body of $f_i$).
	For example, during our evaluation, we observed that 90\% of
	the symbolic predicates $f_i$ have less than 40 lines and
  less than 5 involved input bytes.

	\item Dynamic taint analysis can be utilized to record more
	information than input dependency, including involved operations,
	or even the execute tree.
	Such information can drastically reduce the search space for
	an oracle-guided synthesizer.
\end{enumerate}

Based on these observations, we believe program synthesis could be a
more efficient way to derive $f_i$.

\section{Design}
\label{s:design}


\begin{algorithm}[ht]
\caption{\sys's main loop.}
\label{alg:flip}

\SetKwInOut{Input}{Input}
\Input{$program$: source code of the target program}
\Input{$seeds$: seed inputs}
\BlankLine
$program_{nt}$ $\leftarrow$ compile $program$ w/o taint tracking \\
$program_t$ $\leftarrow$ compile $program$ w/ taint tracking \\
\BlankLine
\ForEach{$\mathbf{x} \in seeds$}{
	\BlankLine
	$trace, labels \leftarrow program_t(\mathbf{x})$ \\
	\BlankLine
	\ForEach{$b \in trace$}{
		\BlankLine
		\tcp{extracts the partial AST of the branch predicate $f_b$ from taint labels}
		\BlankLine
	    $sketch \leftarrow PartialAST(labels, b)$
	    \BlankLine
	    \tcp{extracts the relevant input bytes}
	    \BlankLine
	    $\mathbf{x}_b \leftarrow Args(sketch, \mathbf{x})$
	    \BlankLine
		\tcp{extracted branch output from $trace$}
	    \BlankLine
	    $f_b(\mathbf{x}_b) \leftarrow BranchOut(trace, b)$
	    \BlankLine
		$io \leftarrow \varnothing$ \tcp*{initialize input-output pairs}
                \tcp{initialize new input and branch output to generate}
		$X_b' \leftarrow \mathbf{x}_b$
		$f_b(\mathbf{x}_b') \leftarrow f_b(\mathbf{x}_b)$
		\BlankLine
		\tcp{repeat until the branch output is changed/flipped}
		\While{$f_b(\mathbf{x}_b') = f_b(\mathbf{x}_b)$}{
			\BlankLine
			$io \leftarrow io \cup \{(\mathbf{x}_b',f_b(\mathbf{x}_b')\}$
			\BlankLine
			\tcp{complete the sketch based on existing input/output pairs}
			\BlankLine
			$f_b^{syn} \leftarrow Synth(sketch, io)$
			\BlankLine
			\tcp{get a new assignment of relevant bytes that may flip the branch}
			\BlankLine
			$\mathbf{x}_b' \leftarrow Solve(f_b^{syn}(\mathbf{x}_b') \neq f_b(\mathbf{x}_b))$
			\BlankLine
			\tcp{generate a new input by replacing old relevant bytes with new ones}
			\BlankLine
			$\mathbf{x}' \leftarrow Substitute(\mathbf{x}, \mathbf{x}_b, \mathbf{x}_b')$
			\BlankLine
			\tcp{test the new input}
			\BlankLine
			$trace' \leftarrow program_{nt}(\mathbf{x}')$ \\
			$f_b(\mathbf{x}_b') \leftarrow BranchOut(trace', b)$
		}
                \tcp{Success: $\mathbf{x}_b'$ has new input flipping the target branch}
	}
}

\end{algorithm}

\subsection{Overview}
%
Similar to traditional concolic execution~\cite{ma2011directed,
godefroid2005dart}, for each branch that is \emph{directly} affected by
the input,
\sys aims to generate a new input that flips the evaluation result of
that branch instruction (\autoref{alg:flip}).
More specifically, for target program $P$ which takes a vector of bytes
$\mathbf{x}$ as input and yields an execution trace $T = {b_1,b_2,\dots,b_n}$,
the primary goal of \sys is to construct a new input $\mathbf{x}'$,
which can change the evaluation result of $b_i$ to the opposite
(\ie, $\mathtt{true} \rightarrow \mathtt{false}$ or
$\mathtt{false} \rightarrow \mathtt{true}$).

To achieve this goal, \sys performs the following analyses in order:
\begin{enumerate}
\item A branch-driven {\bf taint analyzer}, which performs a
light-weight data-flow analysis particularly designed to extract
(i) which input bytes in $\mathbf{x}$ are used to evaluate the target branch
and (ii) roughly how these input bytes are used to evaluate the target
branch (\ie, a sketch).

\item A branch predicate {\bf synthesizer}, which uses the taint analysis
result and collected input-output pairs to synthesize the symbolic
branch predicate of the target branch;

\item A branch {\bf flipper}, which uses the synthesis results
(including branch predicates prior to the target branch) to obtains
a new concrete input $\mathbf{x}'$ that may flip the evaluation of the
target branch.
\end{enumerate}

\subsection{Context-Sensitive Branch Tracking}
\label{s:design:context}

As pointed out by Chen~\etal~\cite{angora}, lacking context-sensitivity
could impact a fuzzer's ability to discover interesting cases.
Lacking context sensitivity will cause additional problems for \sys.
In particular, \sys uses two binaries to collect different information:
one binary ($program_t$) with taint analyzer to collect the information
on how input bytes would affect each branch $b_i$ in an execution
trace $T$; and another binary \emph{without} taint analyzer ($program_{nt}$)
to collect input-output pairs for the branch predicate of $b_i$.
We choose this two-binary approach because even though dynamic taint
analysis is faster than traditional concolic execution, it is still
very expensive.
Therefore, we would like to perform taint analysis only when necessary.
For the rest of the time, we would like to use a binary without taint
analysis so we can enjoy a throughput similar to fuzzing.
Moreover, this approach also allows us to instrument $program_{nt}$ with
other error detectors that may not be compatible with the taint analyzer,
such as address sanitizer (ASAN).

The first challenge of this two binary design is how to associate
input-output pairs with $b_i$.
More specifically, the same conditional branch statement $b_{static}$
can appear (be executed) multiple times in the trace $T$; and
their symbolic predicate could be different.
Ideally, we should identify each $b_i$ by its position in the execution
trace $T$;
but as logging and parsing execution trace are expensive,
we opt for AFL's approach---naming each branch with a unique identifier.
Then, to distinguish different execution of the same branch statement,
\sys associates each occurrence of $b_{static}$ with
a \emph{context}:\\
\[
\mathit{ID}(b_i) := context \oplus \mathit{SID}(b_{static})
\]\\
In our current prototype, $\oplus$ is the \cc{xor} operations and
$context$ is defined as the calling context, which is a hash over the
current call stack:\\
\[
context := \mathit{SID}(callsite_1) \oplus \dotsb \oplus \mathit{SID}(callsite_n)
\]\\
While there could be other definition of $context$ that provides less
collision, we choose this approach for its simplicity and the fact that
computing \cc{xor} with the same callsites would have the same effect as ``poping''
the callee from call stack.

$\mathit{SID}()$ is a function that maps an instruction to an unique name.
How to define $\mathit{SID}()$ is the second challenge of the two-binary approach.
In AFL, instructions are identified by its virtual address
(\ie, $\mathit{SID}(inst) := vaddr(inst)$);
however, since \sys uses two binaries,
the same instruction could have different addresses.
An alternative approach, as used in AFL's \cc{llvm_mode}, is to name
each instruction by its numeric order when iterating through instructions.
Unfortunately, two binaries of \sys are built through different
instrumentation passes (\eg, DFSAN for $program_{t}$ or ASAN for $program_{nt}$)
where each instrumentation
pass shows its own numeric ordering, this approach is not reliable
either.
To solve this problem, we prefer to name an instruction by its
location inside the source code (\ie, filename, line number,
and column number):\\
\[
\mathit{SID}(inst) := hash(module(inst) \cdot line(inst) \cdot column(inst)),
\]\\
where $module()$ returns the name of the source file that contain $inst$,
$line()$ and $colum()$ return the line number and column number of the
$inst$, and $\cdot$ is the string concatenation operation.

\subsection{Branch-Driven Taint Analyzer}
\label{s:design:taint}



The goal of a branch-driven taint analyzer is to collect a partial
execution history, which provides guided information (\ie, a sketch)
for the branch predicate synthesizer to reconstruct the symbolic
constraints over $b_i$.

An interesting aspect of our branch-driven taint analyzer is that
in order to scale to complex programs, 
\sys features a unique design positioning somewhere between
traditional dynamic taint analysis and a symbolic execution.
More specifically, traditional dynamic taint analysis aims to answer
data dependencies between sources (\eg, program inputs) and sinks
(\eg, branch predicates); but it cannot answer how a dependency is
formed.
From this perspective, \sys is heavier than traditional dynamic taint
analysis because it collects much more information about the
dependencies, such as operations and common constants.
On the other hand, \sys is much lighter than traditional symbolic
execution because it does not record the complete dependency
information (\ie, in the form the symbolic formula) as
most concrete values are not recorded.
For this reason, \sys does not perform any interpretation---all
instructions are executed on the real hardware at native speed.

Another difference from traditional dynamic taint analysis is that
the sinks of our taint analyzer are conditional branches.
To be more specific, we are interested in the predicate that controls
a conditional branch.
To capture the taint information of a predicate,
\sys instruments comparison related instructions which are
usually ignored in traditional taint analysis.
Given a comparison instruction,
\sys checks if any of the two operands is tainted;
if so, it dumps the taint information of both operands
(concrete operand has a single taint label $untainted$),
as well as the relationship operation (\eg, $=, \leq, >$).

\PP{Byte-Level Taint Tracking}
The foundation of \sys's taint analyzer is a byte-level dynamic
taint analysis from data-flow sanitizer (DFSAN)~\cite{dfsan}.
Dynamic taint analysis extends each memory byte and register with
an additional tag to store the taint label.
A taint analyzer defines three policies~\cite{schwartz2010all}:
(i) how ``taint'' is introduced (\aka, sources),
(ii) when labels are checked (\aka, sinks),
and (iii) most importantly, how labels are propagated,
especially how multiple labels should be merged.
When the label is binary (\ie, the label can only be
\emph{tainted} or \emph{untainted}),
the propagation is relatively simple.
However, supporting byte-level taint tracking
(\ie, each input byte has a unique taint label)
imposes additional challenges on managing labels.
First, a unique taint label is assigned to every input byte of
$\mathbf{x}$.
The number of labels required grows as the input size grows.
Second and more importantly, the number of possible combinations of
the bytes grows in the order of the factorial function.
As a result, it is impossible to store all the information into the
taint label.

To solve this issue, data-flow sanitizer uses a data structure
called \emph{union table} which is indexed by taint labels to store
the actual taint/dependency information.
Labels inside the union table are organized as a forest of binary trees:
all initial input labels are leaf nodes;
and whenever two labels need to be merged
(\eg, involved in a binary operations like \cc{add}),
a new label that points to the two labels as children nodes is created.




\PP{Operation-Aware Taint Analysis}
To reduce the search space of the synthesizer,
the taint analyzer records more information than data dependencies.
This is done by extending the union table.
More specifically,
as the union table from DFSAN is already organized as binary trees where
merging two labels will result in a new table entry with the two input
entries as the child nodes,
\sys extends it to form an abstract syntax tree (AST) like structure.
In the extended union table, each entry is a tuple
$(op, operand_1, operand_2, size)$,
where $op$ is the operation that caused the two labels to merge,
$operand_1$ and $operand_2$ are the corresponding labels that are merged,
and $size$ is the size of the operands.

As a concrete example, let us consider an instruction $c = a + b$.
Because $+$ operation is commutative,
in traditional dynamic taint analysis, there could be three cases:
(i) if neither $a$ nor $b$ is tainted, then $c$ is also not tainted;
(ii) if only one operand is tainted, or they share the same label,
then $c$ would inherit the same label as the tainted one; and
(iii) if both $a$ and $b$ are tainted and their labels are different,
then the union function $U$ would create a new taint label
$l_c := \mathrm{U}(l_a, l_b)$,
where $l_a$ and $l_b$ are taint label for $a$ and $b$, respectively.
In \sys, we only have two corresponding propagation policies.
The first one is the same as traditional taint analysis:
if neither $a$ nor $b$ is tainted, then $c$ is also not tainted.
However, as soon as one of the operand is tainted,
\sys's union function $U_{\scriptsize{\sys}}$ will create a new label
$l_c := U_{\scriptsize{\sys}}(+, l_a, l_b, size)$.

Traditional concolic execution engines also maintain symbolic formula
of a variable as an AST, where the leaf node is either
a symbolic input or a concrete value.
Unlike them, \sys's union function does not record most concrete
values---they all share the same untainted label.
The rationale behind this design decision is two-fold:
(i) the number of unique concrete values can be very large which will quickly
exhaust the taint labels supported by a fixed size union table; and
(ii) most concrete values can be easily recovered by the synthesizer.


%
%
%

\PP{Optimization: Common Constants}
While the taint analyzer does not record most concrete values,
it does record a limited set of common constant values to speed-up
the synthesis process.
Specifically, non-linear operations like bit-wise shifting require
much more input-output pairs to recover the correct operands than
linear operations.
At the same time, many of these non-linear operations frequently use
a small set of concrete operands, \eg, shifting 8, 16, 32 bits.
Based on these observations, besides label $0$, which represents
untainted/concrete value, \sys also reserves the next $n$ labels\footnote{
  In our current prototype, $n = 16$, constants are $8 \times i, i \in [0,15]$.
}
for frequently used constants.

\PP{Optimization: Load and Store}
While our operation-aware byte-level taint analysis already omits
most concrete values to limit the required taint labels,
it could still introduce a larger number of taint labels than
traditional taint analysis.
This negatively impacts \sys in two different aspects:
(i) memory space used by the union table becomes huge and
is exhausted more quickly; and
(ii) it makes the symbolic formula (\ie, $f_i^{syn}$) large and
in many cases, unnecessarily complex.
To address this issue, we designed several optimization techniques.

The first optimization is related to load and store operations.
In traditional concolic execution, both operations work at byte granularity.
As a result, loading data larger than one byte will involve several
``concatenation'' operations;
and storing data larger than one byte will result in several
``extraction'' operations.
For example, considering a simple assignment statement with two 32-bit
integers: $x = y$,
where $x$ and $y$ are to-be-tainted and tainted integers, respectively.
When the load operation is recorded at the byte granularity, \sys needs to
create three new labels:
\begin{itemize}[itemsep=0.1em,label={}]
\item \begin{center} $l_{t1} := U(cat, l_{b1}, l_{b0}, 1)$ \end{center}
\item \begin{center} $l_{t2} := U(cat, l_{b2}, l_{t1}, 1)$ \end{center}
\item \begin{center}    $l_y := U(cat, l_{b3}, l_{t2}, 1)$ \end{center}
\end{itemize}
where $L_{bi}$ represents the taint label of each byte.
To make the matter worse, when storing $L_x$ back to memory, \sys
needs to create additional four labels:
\begin{itemize}[itemsep=0.1em,label={}]
\item \begin{center} $l_{b0}' := U(extract, l_x, 0, 1)$  \end{center}
\item \begin{center} $l_{b1}' := U(extract, l_x, 8, 1)$  \end{center}
\item \begin{center} $l_{b2}' := U(extract, l_x, 16, 1)$ \end{center}
\item \begin{center} $l_{b3}' := U(extract, l_x, 24, 1)$ \end{center}
\end{itemize}

In order to utilize label space more efficiently and makes the symbolic
formula to be synthesized simpler, \sys implements additional
optimizations for load and store operations.
First, \sys uses a special operation $uload$ to express loading
a sequence of bytes: \\
\[
label := U_{\scriptsize{\sys}}(uload, l_{start}, size, size)
\]
\\
where $L_{start}$ represents the label of first byte and
$size$ indicates how many bytes are loaded.
When handling a load operation, \sys will first check if the $uload$
operation is applicable (\ie, the labels of the corresponding bytes
are continuous) before falling back to the old way.
Second, when handling store operations, if the label is a result of
$uload$ operation, \sys will directly extract labels of the
corresponding bytes from the $uload$ operation;
otherwise \sys will assign the same label to all the involved bytes
without breaking them up.
The reason for doing this is because in most cases, integers are
stored and loaded as a whole;
so only when the size of load operation differs from the size of
stored label, \sys will emit an $extract$ operation (\ie, lazy-extract).

\PP{Optimization: Concrete Folding}
Similar to the constant folding technique used by compilers,
\sys also folds concrete values to avoid creating new labels.
That is, when handling binary operations, if one operand is concrete
and the other operand is a result of the same operation;
then we will check if the two concrete values can be combined,
if so, we will use the same label instead of creating a new one.
For example, for statement $x = a + b + c$,
if both $b$ and $c$ are concrete,
we will fold them into a single concrete value.

\PP{Optimization: Avoiding Duplication}
The last optimization is to reuse existing labels so as to reduce
the number of created labels.
This is done through a reverse lookup table.
In particular, the union table maps a label (\ie, index to the table)
to the corresponding taint information $(op, l_1, l_2, size)$.
To avoid creating duplicated taint information,
\sys uses a hash table to map taint information back to its label.
Whenever two labels need to be merged,
\sys first checks whether $(op, l_1, l_2, size)$ already exists.
If so, it reuses the same label; otherwise a new label is allocated to
represent the taint information.
Note that in case $l_1$ or $l_2$ is concrete, this step may introduce
false ``aliasing'' as the actual concrete values could be different.
However, the variables that have same taint labels can have different
values, since concrete values can be different. We mitigate this issue
when synthesizing the branch predicate.

\subsection{Branch Predicate Synthesizer}
\label{s:design:synthesis}


After collecting taint information of branch predicates,
the next step is to reconstruct their symbolic formula.
\sys uses oracle-guided program synthesis~\cite{jha2010oracle} to
achieve this goal.
More specifically, for each branch predicate that can be directly
influenced by the input data, our taint analyzer has already recorded
its partial AST for both sides of the comparison operation
where the concrete operands are missing.
To figure out the correct value assignments to the missing concrete operands,
the synthesizer relies on an SMT solver.
Specifically, by symbolizing all the operands in the partial AST,
including the missing concrete values, the synthesizer first
constructs two symbolic functions (\ie, $f_{b,l}$ and $f_{b,r}$,
representing symbolic functions for left-hand-side and right-hand-side
of the comparison, respectively) with the relevant input bytes as
arguments.
The return value of these functions are the values involved in the
comparison operation.
Then it runs the target program (the version without
dynamic taint analysis) with those relevant inputs bytes mutated to
collect a set of input-output pairs.
By binding the arguments to the concrete input bytes and the return value
to the observed concrete operands of the comparison operation,
the synthesizer creates a set of constraints over the missing
concrete operands.
Finally, it consults the SMT solver for a feasible assignment of those
concrete operands.

If the solver can find a feasible assignment---this is not always
possible, we will discuss later why this can happen---we have
a candidate predicate formula.
The next step is to consult the SMT solver again for a new input
that can flip the branch.
This is done by removing the previous binding of the arguments and
return values and bind the concrete values to the feasible assignment.
With $f_{b,l}$ and $f_{b,r}$ ready, the synthesizer can then ask the
SMT solver for a new set of relevant inputs bytes such that the
evaluation of the comparison operation would be negated.
However, because the synthesized functions can be wrong,
the new input may not always be able to flip the branch.
In this case, we have a \emph{counterexample}.
To fix the incorrect functions, we add this new input-output pair to
the set to query for a new assignment.
This is repeated until we found the correct formula or
the number of iterations have reached a certain limit
(\eg, 10 in our prototype).
Next, we will use illustrative example (\autoref{fig:synthexample}) to
show how this process works.

\begin{figure}[t]
  \begin{subfigure}{0.45\textwidth}
    \input{code/orig.c}
    \caption{Original code of target branch.}
    \label{code:orig}
  \end{subfigure}
  \begin{subfigure}{0.45\textwidth}
  	\begin{scriptsize}
    \begin{itemize}[label={},leftmargin=*]
	  \item $l_{81} := U_{\scriptsize{\sys}}(or, l_{79},l_{80}, 1)$
	  \item $l_{80} := U_{\scriptsize{\sys}}(ZeroExt, l_{76}, 4, 1)$
	  \item $l_{79} := U_{\scriptsize{\sys}}(<<, l_{78}, 0, 4)$
	  \item $l_{78} := U_{\scriptsize{\sys}}(ZeroExt, l_{77}, 4, 1)$
	  \item $l_{77} := U_{\scriptsize{\sys}}(and, 0, l_{17}, 1)$
	  \item $l_{76} := U_{\scriptsize{\sys}}(and, 0, l_{18}, 1)$
	\end{itemize}
	\end{scriptsize}
	\caption{Logged taint labels relevant to the branch.}
	\label{code:uniont}
  \end{subfigure}
  \begin{subfigure}{0.45\textwidth}
    \input{code/synth.py}
    \caption{Synthesized function for the left-hand-side of the comparison.}
    \label{code:synth}
  \end{subfigure}
  \begin{subfigure}{0.45\textwidth}
    \input{code/solve.py}
    \caption{How input-output pairs are used to construct constraints
    and get feasible assignments to the concrete operands.}
    \label{code:solve}
  \end{subfigure}
  \caption{An example of reconstructing symbolic expressions}
  \label{fig:synthexample}
  \vspace{-2em}
\end{figure}

\PP{Reconstructing Symbolic Predicate}
When we execute the code in~\autoref{code:orig},
let us assume the taint analyzer logged a branch predicate
$bp_3 = (l_{81}, 254, equal, 4)$,
where $bp_3$ means a conditional branch at Line~3,
$l_{81}$ means the left-hand-side of the comparison (\ie, variable \cc{c})
is a tainted value stored in the entry 81 of the taint label table,
$254$ means the right-hand-side is a concrete value \cc{254},
$equal$ means the relationship operations is \cc{==},
and $4$ means the operands size ($l_{81}$ and $254$) are 4 bytes.
The relevant taint labels (\ie, the partial AST for \cc{c}) are in
\autoref{code:uniont}.

\sys synthesizes its symbolic formula as follows.
First, for each side of the comparison operation that is tainted
the synthesizer creates a symbolic function: $f_{i,l}$ for
left-hand-side (if tainted) and $f_{i,r}$ for right-hand-side (if tainted).
In the above example, only the left-hand-side (\ie, variable \cc{c})
is tainted, so we only need to synthesize the formula for \cc{c}.
Next, to populate the parameters and the body of $f_{i,\{l,r\}}$,
the synthesizer parses the serialized partial AST from leaves to the root.
Each leaf node (\ie, label belongs to the raw input bytes)
is added as an argument to the function.
For each non-leaf node, a corresponding statement is added to the
function's body where symbolic variables are created for both tainted
and untainted labels.
Considering the example above whose partial AST is shown
in~\autoref{code:uniont}.
The leave nodes are $l_{17}$ (variable \cc{utf[0]}) and
$l_{18}$ (variable \cc{utf[1]}).
For each non-leaf node (from $l_{76}$ to $l_{81}$), a line is added,
based on the involved operations and the operands.
If an operand has label $l_0$ (\ie, a concrete value), we create a new
symbolic variable 
based on the operand size (\eg, a bitvector variable \cc{c0} with 8 bits).
Finally, the root node of the partial AST will be used as the return
value of the function.
\autoref{code:synth}~\footnote{\cc{BitVec(n, w)} returns a \cc{w}-bits
  bitvector, which can be referenced using a name
  {n}. \cc{BitVecVal(v, w)} returns a \cc{w}-bits bitvector, which is
  initialized with a concrete value \cc{v}.} shows the constructed
function for the left-hand-side of the branch predicate
in~\autoref{code:orig}.  Note that since our prototype uses Z3-python
binding, the types of variables $v76$ to $v81$ are automatically
inferred.

\PP{Collecting Input-Output Pairs}
After reconstructing the partial symbolic formula,
the next step is to collect input-output pairs for solving the concrete
values inside the symbolic formula.
This is done by (i) mutating the argument bytes of $f_{i,\{l,r\}}$
(\eg, in the above example, the 16th byte (\cc{v17}) and
the 17th byte (\cc{v18}))
(ii) execute the target program with mutated inputs, and
(iii) collect the operands of the comparison operation
(\ie, the output of $f_{i,\{l,r\}}$).
Note that because the program we run in this step is \emph{without}
dynamic taint analysis, it executes at the native fuzzing speed.
The output is logged to a file in the following format:\\
\[
(\mathit{ID}(b_i), o_{i,l}, o_{i,r}, o_i)
\]\\
where $\mathit{ID}(b_i)$ is the identifier of the target branch
(as defined in \autoref{s:design:context}),
$o_{i,\{l,r\}}$ are the concrete values of the operands in the
comparison operation, and $o_i$ is the result of the comparison
operation (\ie, the value of the branch predicate,
either $\mathtt{true}$ or $\mathtt{false}$).

\PP{Solving Concrete Values}
To find out the correct assignment of the concrete values inside the
symbolic formula, we use the SMT solver.
In particular, we first treat all concrete values as symbolic and
input-output values as concrete.
And for each input-output pair, we add a constraint:\\
\[
assert(f_{i,\{l,r\}}(\mathbf{x}_{i,\{l,r\}}) = o_{i,\{l,r\}})
\]
\\
where $\mathbf{x}_{i,\{l,r\}}$ is the input arguments to the corresponding
function we reconstructed in step one and $o_{i,\{l,r\}}$ is the
output we collected in step two.
After adding such constraints for all input-output pairs,
the synthesizer then asks the SMT solver to check the satisfiability.
If the constraints are satisfiable, we can get an assignment to all
concrete values from the solver.
For example, to solve \cc{c0,c1,c2} in~\autoref{code:synth},
we could add the constraints in~\autoref{code:solve}.

\PP{Optimistic Solving}
Although we have added context sensitivity in our branch tracking,
it is still not as accurate as its path-sensitive counterpart.
Together with hash collision~\cite{gan2018collafl},
they introduced a problem for solving the concrete values.
Specifically, some of the input-output pairs we collected in step two
could be wrong.
As a result, we could impose incorrect constraints thus prevents the
SMT solver from finding the correct assignment.
To mitigate the impact of this problem,
we perform \emph{optimistic} solving.
That is, instead of blindly adding all input-output pairs as constraints,
\sys first checks if the new constraint to be added is compatible with
existing model (\ie, adding it will not make the model unsatisfiable).
If so, the constraint will be added; otherwise it will be dropped.

\subsection{Branch Flipper}

After synthesizing both sides of the predicate for the target branch,
the next step is to construct a new input that could flip the branch,
\ie, forcing the execution to go through the opposite path.
This goal is to achieve similarity to traditional concolic execution engines.

First, \sys constructs the symbolic formula of the branch predicate.
After getting a feasible assignment of concrete values for both
operands of the comparison operation,
\sys ``concretize'' them by replacing them with the corresponding
concrete values.
For example, in~\autoref{fig:synthexample}, we re-assign \cc{c0,c1,c2}
to the corresponding concrete values:

\begin{figure}[ht]
\input{code/constants.py}
\end{figure}

Now we have two completed symbolic functions $f_{i,l}(\mathbf{x}_{i,l})$
and $f_{i,r}(\mathbf{x}_{i,r})$ and the predicate can be easily
constructed as:
\vspace{-0.3em}

\begin{equation}
\label{f:predicate}
f_i := f_{i,l}(\mathbf{x}_{i,l}):relop:f_{i,r}(\mathbf{x}_{i,r})
\end{equation}

where $relop$ is the relationship operation (\eg, $=, \leq, >$).
Next, \sys parses the execution log to find out the value(s) ($o_i$)
of the predicate.
Then we consult the SMT solver for the satisfiability of the following
constraint:
\vspace{-0.3em}

\begin{equation}
\label{f:query}
assert(f_i = \neg o_i)
\end{equation}

If the constraint is satisfiable (\ie, the opposite branch is feasible),
\sys then gets a feasible assignment to the input bytes
($\mathbf{x}_{i,r}$ and $\mathbf{x}_{i,l}$) and constructs a new input by replacing
the values of the those bytes with the assignments from the SMT solver.
However, compare to traditional concolic execution,
there are two additional challenges in \sys.

\PP{Incorrect Synthesis Result}
Unlike traditional concolic execution, the symbolic expression
generated by the synthesizer may not be correct.
This is because there could be multiple feasible assignments to the
concrete values that potentially satisfy the constraints derived from the
input-output pairs,
especially when non-linear operations are involved.
As a result, the assignment we use in~\autoref{f:predicate} may not be
the correct one.
To fix this problem, \sys will execute the target program with the
new input generated by the branch flipper and watch whether the target
branch $b_i$ has been flipped (\ie, $o_{i}' \neq o_i$).
If not, this input and the logged output ($o_{i,l}$ and $o_{i,r}$)
will be added back to the input-output pair to get new assignment of
the concrete values and the branch flipper will try to generate a new
input based on the new assignment.
This loop is repeated until the branch is flipped
(\ie, we have synthesized the correct symbolic expression) or
the number of iteration has reached a threshold (\eg, 10 rounds),
as we do not want to spend a lot of time on a particular branch.

\PP{Nested Conditions}
A limitation of previous taint-guided fuzzers~\cite{vuzzer,angora} is
that they only consider one branch at a time.
However, the input bytes that can affect the target branch could also
affect other branches, especially branches that are executed before
the target branch.
As a result, the derived input that aims to flip the target branch
could accidentally flip a preceding branch thus completely divert the
execution.
The ability to recover the correct symbolic expression of a branch
predicate enables \sys to overcome this limitation.
In particular, when capturing the taint information,
the taint analyzer logs branches in the chronological order they are executed.
During synthesis, the synthesizer also processes branches in the same
order, which means when trying to flip a branch,
we should already have the symbolic expressions of all the branch
predicates before this branch.
To avoid accidental flips of an earlier branch,
the branch flipper maintains a global context for each executed trace
that is logged by the taint analyzer where all the symbolic functions
$f_{i,\{l,r\}}$ are visible and all input bytes $\mathbf{x}$ are \emph{global}.
Then, after successfully generating a new input that satisfies~\autoref{f:query},
the branch flipper will impose the following constraint to force the
execution to follow the same path:
\vspace{-0.3em}

\begin{equation}
\label{f:force}
assert(f_i = o_i)
\end{equation}

Since all input bytes (\eg, \cc{v17,v18}) are global,
when checking the satisfiability of~\autoref{f:query},
the SMT solver will also consider all previous predicates where\\
\[
(\mathbf{x}_{j,l} \cup \mathbf{x}_{j,r}) \cap (\mathbf{x}_{i,l} \cup \mathbf{x}_{i,r}) \neq \emptyset,\quad 0 \leq j < i
\]\\
The drawback, however, is that this will make the solver take longer
time to check~\autoref{f:query}.
In~\autoref{s:eval}, we will compare the performance of \sys when
enabling and disable this ability, in terms of code coverage and speed.

\section{Implementation}
\label{s:impl}

In this section, we present some implementation details of our prototype.

\subsection{Taint Analyzer}

Our dynamic taint analyzer is implemented based on the DataFlowSanitizer
(DFSAN)~\cite{dfsan},
which is part of the LLVM compiler toolchain.
It instruments the source code of the target program with taint
propagation logics.

\begin{table}[ht]
  \centering
  \caption{
    Memory layout of the program for taint analysis.
  }
  \label{table:layout}
  \begin{tabular}{llll}
  \toprule
  \textbf{Start} & \textbf{End} & \textbf{Description} \\
  \midrule
  0x700000020000 & 0x800000000000 & application memory \\
  0x400010000000 & 0x700000020000 & hash table         \\
  0x400000000000 & 0x400010000000 & union table        \\
  0x000000020000 & 0x400000000000 & shadow memory      \\
  0x000000000000 & 0x000000010000 & reserved by kernel \\
  \bottomrule
\end{tabular}

  \vspace{-1em}
\end{table}

\PP{Taint Storage}
Since the instrumentation is actually performed at the LLVM
intermediate representation (IR) level, instead of tagging registers,
DFSAN inserts shadow variables to store tags for all local variables,
including arguments and return values.
Tags of heap variables are accessed via shadow memory.
\autoref{table:layout} shows the memory layout of an instrumented
program:

\begin{itemize}
\item \cc{shadow memory} maps each byte of allocated heap memory to
    its taint label (\ie, index to the union table);
\item \cc{union table} stores the actual taint information, each
    entry stores the tuple $(op, l_1, l_2, size)$ where $l_1$ and
    $l_2$ are also taint labels.
\item \cc{hash table} maps a tuple $(op, l_1, l_2, size)$ to its
    taint label, as discussed in~\autoref{s:design}, this is a new
    data structure introduced by \sys to avoid duplicated labels.
\item \cc{application memory} is the memory used by the target application.
\end{itemize}

To enforce this memory layout, a linker script is used to restrict
the application memory range to avoid colliding with other designated
regions.
Then, once the program starts, the runtime library of the taint analyzer
reserves the designated regions, so the OS kernel will not allocate
virtual addresses within these regions to the application.

\PP{Taint Sources}
To assign labels to input bytes, \sys instruments file related functions.
In our current prototype, we only support tainting data from file and \cc{stdin};
tainting data from network is not supported yet but can be easily extended.
When the program opens a file which should be tainted,
\sys calculates the size of the file and reserves the input label entries.
When the program reads from the file, \sys calculates the offset (within
the file) and the size of the bytes to be read, and assigns the
corresponding labels to the target buffer which receives the read bytes.

Occasionally, the target program may try to read data over the end of
the file, \ie, the bytes to be read is larger than available data.
In this case, \sys tags the remaining buffer with label \cc{-1} and
``deceive'' the program as if the data is available.
Later, if such data has actually been used, \eg, to calculate the
value of a branch predicate, this special label will logged to
indicate that the target program can handle larger input data.
Finally, when this special label is found in the taint log,
\sys will increase the input data size and re-run the target program.

\PP{Taint Propagation}
Our taint propagation policies are almost identical to DFSAN,
the only difference is that when combining two labels,
we will also record the operation that causes the union,
and the size of the operation.
The following (bitvector) operations are supported:

\begin{itemize}
	\item Bit-wise operations: \cc{bvnot}, \cc{bvand}, \cc{bvor},
	\cc{bvxor}, \cc{bvshl}, \cc{bvlshr}, \cc{bvashr};
	\item Arithmetic operations: \cc{bvneg}, \cc{bvadd}, \cc{bvsub},
	\cc{bvmul}, \cc{bvudiv}, \cc{bvsdiv}, \cc{bvurem}, \cc{bvsrem};
	\item Truncation and extension: \cc{bvtrunc}, \cc{bvzext},
	\cc{bvsext};
	\item Special load (as defined in~\autoref{s:design:taint}):
	\cc{bvload}, \cc{bvextract}.
\end{itemize}

In our current prototype, we do not propagate taint based on control
dependencies (\aka, indirect taint) and floating point operations are
also not supported yet.

\PP{Taint Sinks}
In our current prototype, we consider \cc{icmp} and \cc{switch}
instructions as taint sinks (\ie, coverage-oriented).
For \cc{icmp} instruction, \sys checks whether any its operands
are tainted; if so, it serialize the taint information.
First, it dumps the predicate it self $(l_{lhs}, l_{rhs}, op, size)$.
Then for $l_{lhs}$ and $l_{rhs}$, \sys recursively dumps the taint
information (\ie, the partial AST).
For \cc{switch} instruction, \sys treats each case as a comparison
between the condition variable and the case value;
and dumps the taint information.

\subsection{Branch Tracking}

Branch tracking is implemented as a separate LLVM pass which is based
on the instrumentation pass of AFL's \cc{llvm_mode}.
Besides the tracking of control transfer between basic blocks,
the following logics are added.
Firstly, \sys assigns an unique id for each \cc{icmp} instruction and
each case of the \cc{switch} instructions.
As discussed in~\autoref{s:design:context},
the id is calculated using the module name, the line number, and
the source code number of the corresponding instruction, if available.
A module level id list is maintained to detect collisions.
When a collision happens (\eg, for \cc{switch} instruction),
we will append the concrete value involved in the comparison to resolve
the collision.
For the hash function, we used \cc{std::hash<std::string>()}.
\sys also inserts a call to a runtime function to dump the output of
the comparison (during synthesizing) in the format:\\
\[
(\mathit{ID}(\mathtt{icmp}), V_{lhs}, V_{rhs}, V_{icmp})
\]

Secondly, \sys inserts a new global variable to store the $context$
information.
To update the $context$,
\sys first assigns each \cc{call} instruction an unique id using
the same formula as described above.
Then it inserts an \cc{xor} operation before the \cc{call}
instruction to updated the \cc{context} variable with $\mathit{ID}(\mathtt{call})$.
Finally, another \cc{xor} operation with $\mathit{ID}(\mathtt{call})$ is
inserted \emph{after} the \cc{call} instruction to ``pop'' the callee.

\subsection{Synthesizer}

The synthesizer is implemented using Python for its ability to
dynamically create functions through \cc{exec()} and good support from the Z3~\cite{de2008z3}.

\section{Hybrid Fuzzer}
\label{s:fuzzer}


To evaluate \sys, we also implemented a hybrid fuzzer based on \sys
and AFLFast~\cite{bohme2016coverage}.
The hybrid fuzzer takes two binaries, $program_{nt}$ for normal
execution, including fuzzing and input-output pairs collection;
and $program_t$ for taint analysis.
%
Overall, our hybrid fuzzer follows the same cross-seeding design as
previous hybrid fuzzers~\cite{ma2011directed,godefroid2005dart,
stephens2016driller,yun:qsym}.
Specifically, whenever the fuzzer schedules a seed (\ie, input that
provides new coverage) for mutating,
if the seed is not skipped and has not been executed by the concolic
execution engine, it is fed to \sys.
Having received this new input, \sys first collects the branch trace
$T$ by executing $program_t$ with the seed input.
Next, for each branch in $T$, \sys first mutates the relevant
inputs bytes to prepare a initial set of input-output pairs;
then it invokes the synthesizer to create a new input that may flip
the target branch;
if this new input indeed flips the target branch,
\sys moves on to the next branch;
otherwise it adds the branch output to the i/o pairs and asks the
synthesizer to create a new input, until a certain limit is reached.
Because \sys and the fuzzer share the same method to execute
$program_{nt}$, if the input generated by the synthesizer is able to
trigger new coverage, it will automatically be added to the seeds.
Sharing the same execution method also allows \sys to utilize
\cc{fork_server} and \cc{persistent_mode} to speed up the execution.

\PP{Parallel Fuzzing.}
To effectively find target branch predicates to be synthesized,
we have implemented parallel fuzzing which can synchronize from and to
AFL fuzzers. Similar to AFL's parallel fuzzing, \sys automatically
fetches new inputs added to AFL fuzzers, and tries to synthesize
branches that the new inputs generated. If \sys is able to generate
the inputs that can flip branch predicates, \sys checks whether
the inputs cover new paths by checking differences in bitmaps of
AFL fuzzer. If they cover new paths, they are added to
own queue of \sys and automatically synchronized to AFL's queue by
AFL fuzzers.

\section{Evaluation}
\label{s:eval}

In this section, we evaluate the performance of \sys with
a set of real-world softwares and two standard benchmarks:
LAVA-M~\cite{dolan2016lava} and
Google Fuzzer Test Suite (FTS)~\cite{FTS}.
Note that Google FTS also consists of real-world libraries from the
OSS-Fuzz project, such as \cc{freetype}, \cc{openssl};
and bugs are not injected.
Our evaluation focused on the following aspects:

\begin{itemize}
	\item {\bf Scalability.} The design goal of \sys is to make
	concolic execution scalable to real-world softwares;
	so the first question we would like to answer is how
  effective is the synthesis-based approach in term of
  code coverage (\autoref{s:eval:scale}).

	\item {\bf Bug finding.} For programs with known vulnerabilities,
	we would like to know whether \sys can improve the number of found
  bugs and the speed of finding bugs (\autoref{s:eval:bugs}).

  \item {\bf Synthesis.} The main feature of \sys is to synthesize
  branch conditions so that we would like to know how many
  branch conditions can be correctly synthesized and flipped
  (\autoref{s:eval:synthesis}).

	\item {\bf Multi-branch solving.} One feature of \sys is its
	capability to flip a branch without affecting previous branches;
	so we would like to know how important this feature is for testing
	real-world applications and study the cases (\autoref{s:eval:nested}).

\end{itemize}

%
%
%
%

\begin{table*}[t]
  \center
  \caption{Code coverage of real-world applications.}

\begin{tabular}{lcrrrrrrrrrr}
	\toprule
 	\multirow{2}{*}{App}  &   Type      &  AFLFast & LAF & QSYM & \multicolumn{7}{c}{SynFuzz} \\   \cmidrule{2-12}
                        & (C/C++)     &  Avg & Avg  & Avg  & Avg  & vs. AFLFast & p-value & vs. LAF & p-value & vs. QSYM & p-value \\   \midrule
   binutils             &    C        & 3082 & 3100 & 3118 & 3187 &  3.41\% &  1.80E-04   &  2.81\%  &  1.81E-04    &  2.21\% &  1.80E-04 \\
   file                 &    C        & 3795 & 3801 & 4361 & 4406 & 16.10\% &  1.80E-04   & 15.92\%  &  1.80E-04    &  1.03\% &  1.76E-04 \\
   libjpeg              &    C        & 5121 & 5511 & 5493 & 5706 & 11.45\% &  1.77E-04   &  3.54\%  &  1.81E-04    &  3.88\% &  1.81E-04 \\
   libpng               &    C        & 7567 & 7577 & 7643 & 7767 &  2.63\% &  1.77E-04   &  2.50\%  &  1.81E-04    &  1.62\% &  2.09E-04 \\
   libtiff              &    C        & 9262 & 9771 &11778 &13581 & 46.62\% &  1.81E-04   & 38.99\%  &  1.81E-04    & 15.31\% &  1.82E-04 \\
   tcpdump              &    C        &19334 &20651 &22085 &22800 & 17.93\% &  1.82E-04   & 10.41\%  &  1.82E-04    &  3.24\% &  1.81E-04 \\
  \bottomrule
\end{tabular}

  \label{table:realworld}
  \vspace{-1em}
\end{table*}

\begin{table*}[t]
  \center
  \caption{Code coverage of Google FTS. Increases are all measured
  against AFLFast.
  Note that {\bf bold} means the result is statistically significant,
  it does {\em not} mean the result is best.}
  \begin{tabular}{ll r r r r r r r r r r r}
	\toprule
	\multirow{2}{*}{App} & \multirow{2}{*}{\begin{tabular}[c]{@{}c@{}}Type\\ (C/C++)\end{tabular}} &
	\multirow{2}{*}{Total} & AFLFast &  \multicolumn{2}{c}{\begin{tabular}[c]{@{}c@{}}QSYM\end{tabular}} &
  \multicolumn{3}{c}{SynFuzz w/o multi-branch} &
  \multicolumn{3}{c}{SynFuzz} \\ \cmidrule{4-12}
	                            &       &        &   Avg  &   Avg &Inc(\%)&  Avg   & Inc(\%)       &           p-value &   Avg & Inc(\%)        & p-value  \\\midrule
	boringssl-2016-02-12        & C     &  47473 &  5099  &  5099 &  0.00 &  5099  &          0.00 &          2.00E+00 &  5099 &           0.00 &          2.00E+00 \\
	c-ares-CVE-2016-5180        & C     &    179 &   115  &   115 &  0.00 &   115  &          0.00 &          2.00E+00 &   115 &           0.00 &          2.00E+00 \\
	freetype2-2017              & C     &  74447 & 27852  & 31345 & 12.54 & 29906  &          7.37 &          8.83E-02 & 34209 & \textbf{22.82} & \textbf{1.36E-04} \\
	guetzli-2017-3-30           & C++   &   5155 &  3725  &  4178 & 12.15 &  4125  &         10.74 &          3.36E-01 &  4125 &          10.74 &          3.38E-01 \\
	harfbuzz-1.3.2              & C++   &  16279 &  9815  & 10144 &  3.35 &  9893  &          0.79 &          2.25E-01 & 10372 &  \textbf{5.67} & \textbf{1.09E-04} \\
	json-2017-02-12             & C++   &   2569 &  2243  &  2243 &  0.00 &  2243  &          0.00 &          1.03E+00 &  2243 &           0.00 &          1.03E+00 \\
	lcms-2017-03-21             & C     &  21466 &  4139  &  4902 & 18.42 &  4483  &          8.30 &          5.70E-01 &  5316 & \textbf{28.43} & \textbf{1.81E-04} \\
	libarchive-2017-01-04       & C     &  36280 &  6733  &  6800 &  1.00 &  6810  &          1.15 &          6.73E-01 &  6790 &  \textbf{0.85} & \textbf{2.37E-02} \\
	libjpeg-turbo-07-2017       & C     &  21873 &  5279  &  5707 &  8.12 &  5595  &          5.99 &          4.50E-01 &  6016 & \textbf{13.96} & \textbf{8.37E-05} \\
	libpng-1.2.56               & C     &  10848 &  3008  &  3163 &  5.14 &  3263  & \textbf{8.49} & \textbf{1.44E-04} &  3320 & \textbf{10.36} & \textbf{1.45E-04} \\
	libssh-2017-1272            & C     &  23675 &  2085  &  2085 &  0.00 &  2085  &          0.00 &          2.00E+00 &  2085 &           0.00 &          2.00E+00 \\
	libxml2-v2.9.2              & C     & 114842 & 10885  & 10763 & -1.12 & 10824  &         -0.56 &          1.85E-01 & 10824 &          -0.56 &          6.36E-02 \\
	llvm-libcxxabi-2017-01-27   & C++   &   4075 &  3660  &  3660 &  0.00 &  3673  & \textbf{0.36} & \textbf{1.36E-02} &  3673 &  \textbf{0.36} & \textbf{1.36E-02} \\
	openssl-1.0.1f              & C     & 102502 & 12863  & 13944 &  8.79 & 16306  &         26.77 &          2.41E-01 & 13694 &           6.46 &          5.71E-01 \\
	openssl-1.0.2d              & C     &  10801 &  3624  &  3722 &  2.71 &  3627  & \textbf{0.10} & \textbf{1.61E-03} &  3627 &  \textbf{0.10} & \textbf{1.61E-03} \\
	openssl-1.1.0c-bignum       & C     &  89265 &  3416  &  3416 &  0.00 &  3416  &          0.00 &          2.00E+00 &  3522 &  \textbf{3.10} & \textbf{1.82E-03} \\
	openssl-1.1.0c-x509         & C     &  89212 & 13462  & 13462 &  0.00 & 13462  &          0.00 &          2.00E+00 & 13462 &           0.00 &          2.00E+00 \\
	openthread-2018-02-27-ip6   & C++   &  49640 &  4841  &  4901 &  1.24 &  5013  & \textbf{3.55} & \textbf{9.14E-05} &  5013 &  \textbf{3.55} & \textbf{4.04E-05} \\
	openthread-2018-02-27-radio & C++   &  49507 & 10015  & 10368 &  3.52 & 10106  & \textbf{0.91} & \textbf{1.08E-02} & 10313 &  \textbf{2.98} & \textbf{1.09E-04} \\
	pcre2-10.00                 & C     &  19852 & 13940  & 14242 &  2.17 & 14335  & \textbf{2.83} & \textbf{1.95E-02} & 14252 &  \textbf{2.24} & \textbf{1.28E-04} \\
	proj4-2017-08-14            & C     &  21004 &  4918  &  5418 & 10.16 &  5700  &         15.90 &          5.96E-01 &  5589 &          13.64 &          2.73E-01 \\
	re2-2014-12-09              & C++   &   8986 &  6007  &  5938 & -1.14 &  6020  &          0.22 &          3.68E-01 &  6020 &           0.22 &          3.68E-01 \\
	sqlite-2016-11-14           & C++   &  70177 &  5143  &  5167 &  0.46 &  5408  & \textbf{5.15} & \textbf{3.38E-02} &  5017 &          -2.46 &          4.71E-01 \\
	vorbis-2017-12-11           & C     &  11517 &  3658  &  3845 &  5.12 &  3810  & \textbf{4.17} & \textbf{4.13E-05} &  3810 &  \textbf{4.17} & \textbf{4.13E-05} \\
	woff2-2016-05-06            & C++   &  12812 &  3038  &  3061 &  0.75 &  3122  &          2.76 &          5.08E-02 &  2985 & \textbf{-1.74} & \textbf{8.68E-05} \\
	wpantund-2018-02-27         & C     &  19976 &  7345  &  7734 &  2.42 &  7634  & \textbf{3.93} & \textbf{1.40E-04} &  7261 & \textbf{-1.16} & \textbf{1.54E-02} \\
  \bottomrule
\end{tabular}

  \label{table:fts}
  \vspace{-1em}
\end{table*}

\begin{figure}[t]
  \centering
  \begin{subfigure}[t]{0.23\textwidth}
    \includegraphics[width=\textwidth]{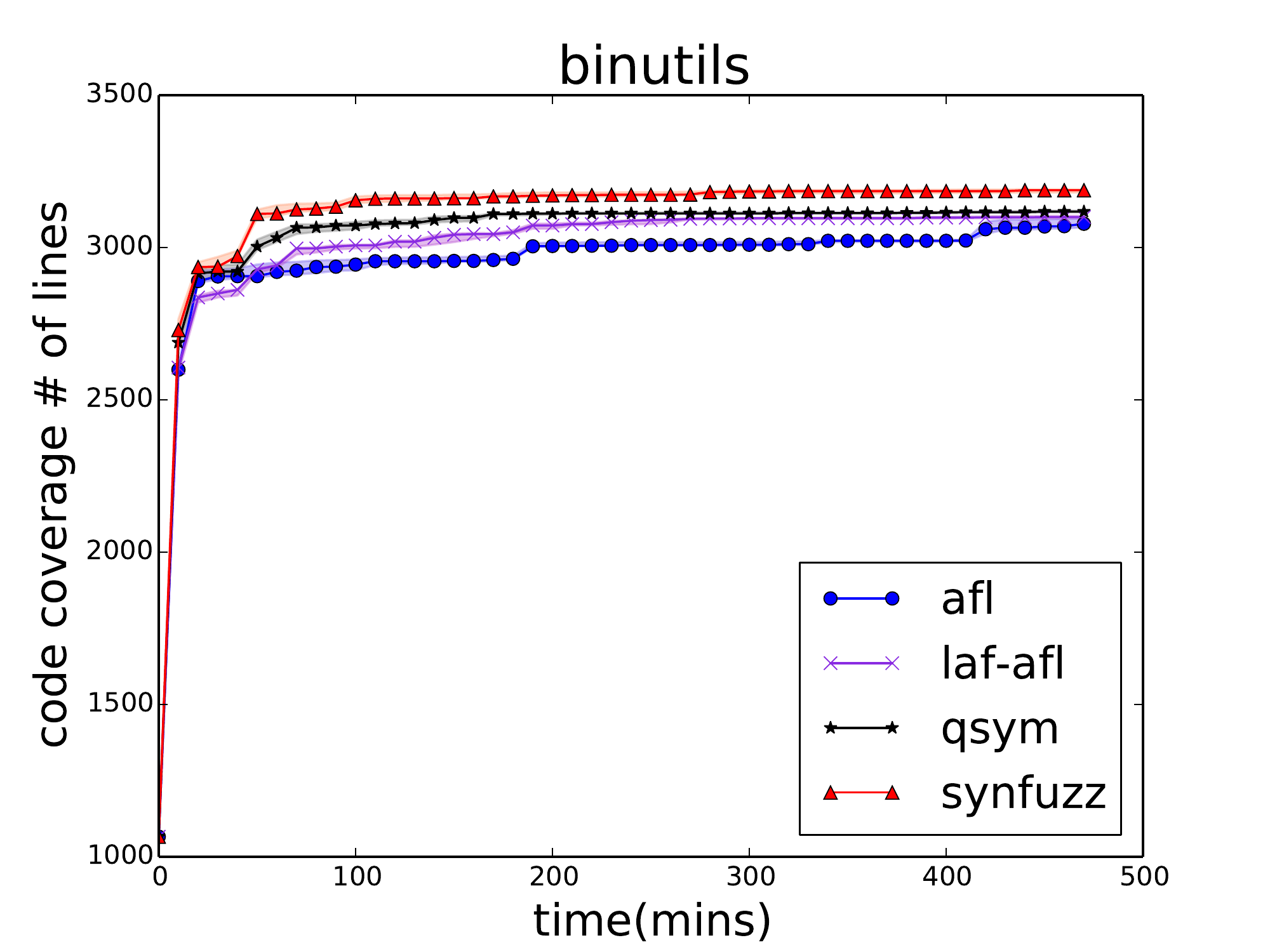}
  \end{subfigure}
  \begin{subfigure}[t]{0.23\textwidth}
    \includegraphics[width=\textwidth]{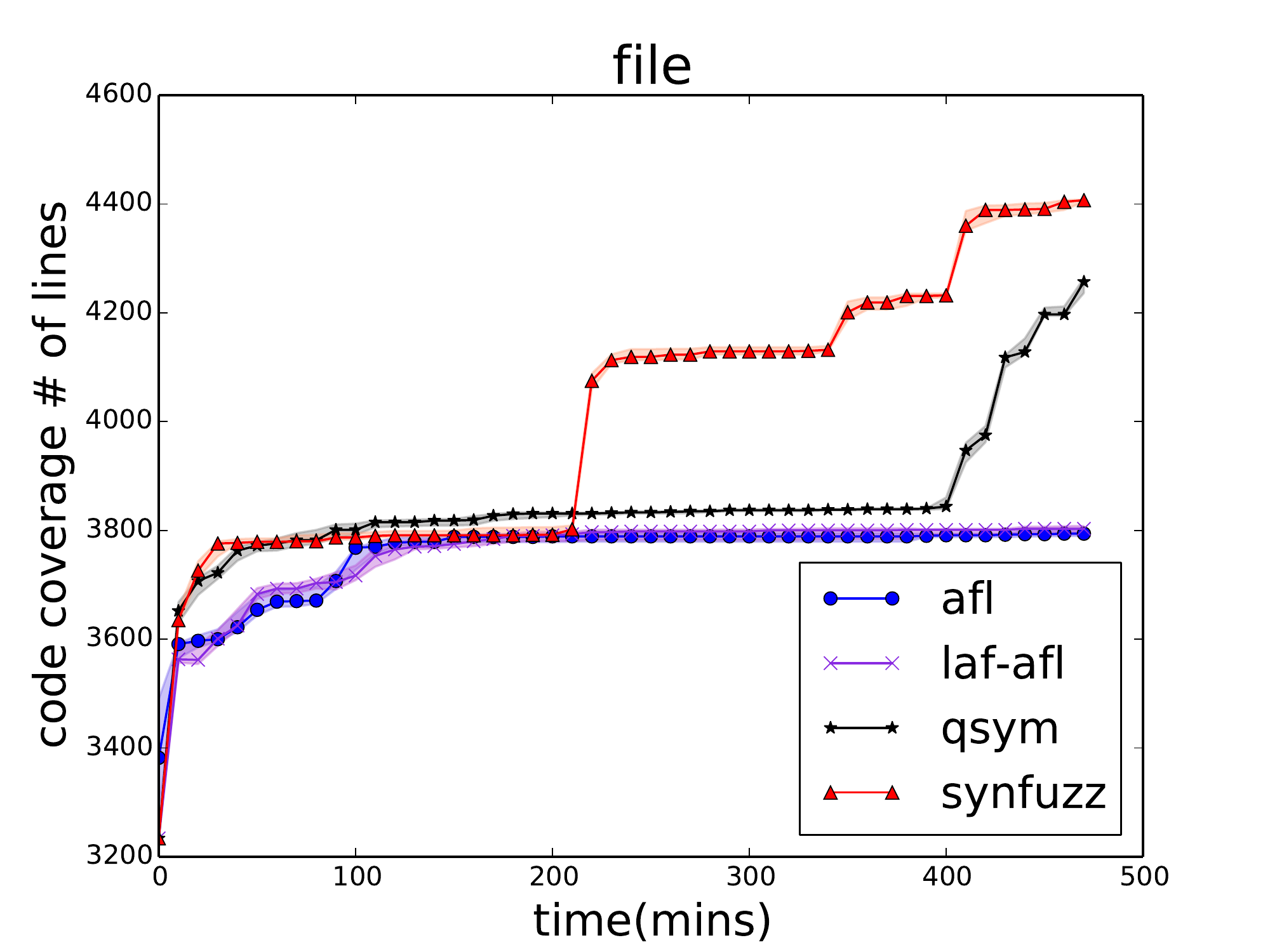}
  \end{subfigure}
  \begin{subfigure}[t]{0.23\textwidth}
    \includegraphics[width=\textwidth]{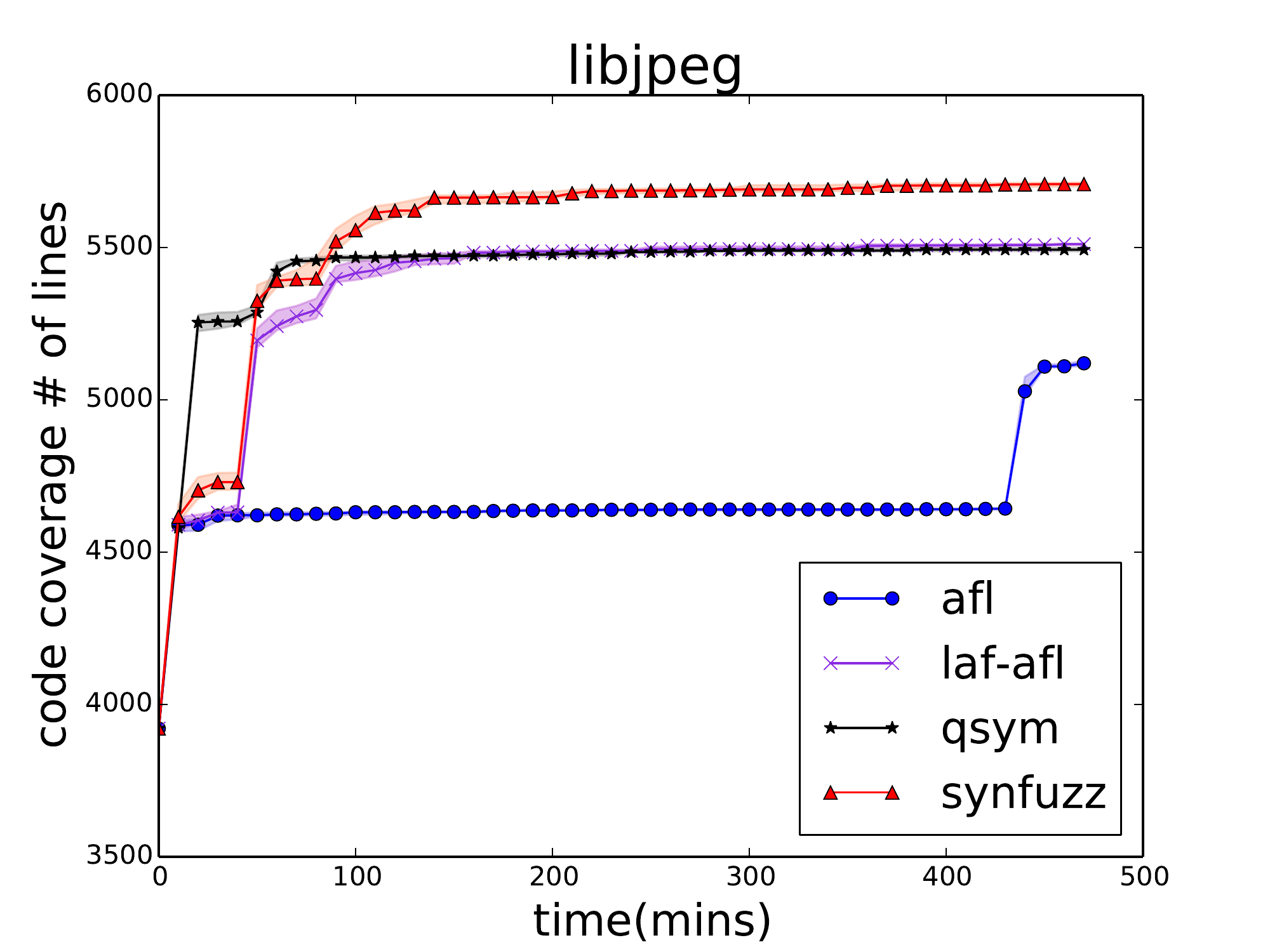}
  \end{subfigure}
  \begin{subfigure}[t]{0.23\textwidth}
    \includegraphics[width=\textwidth]{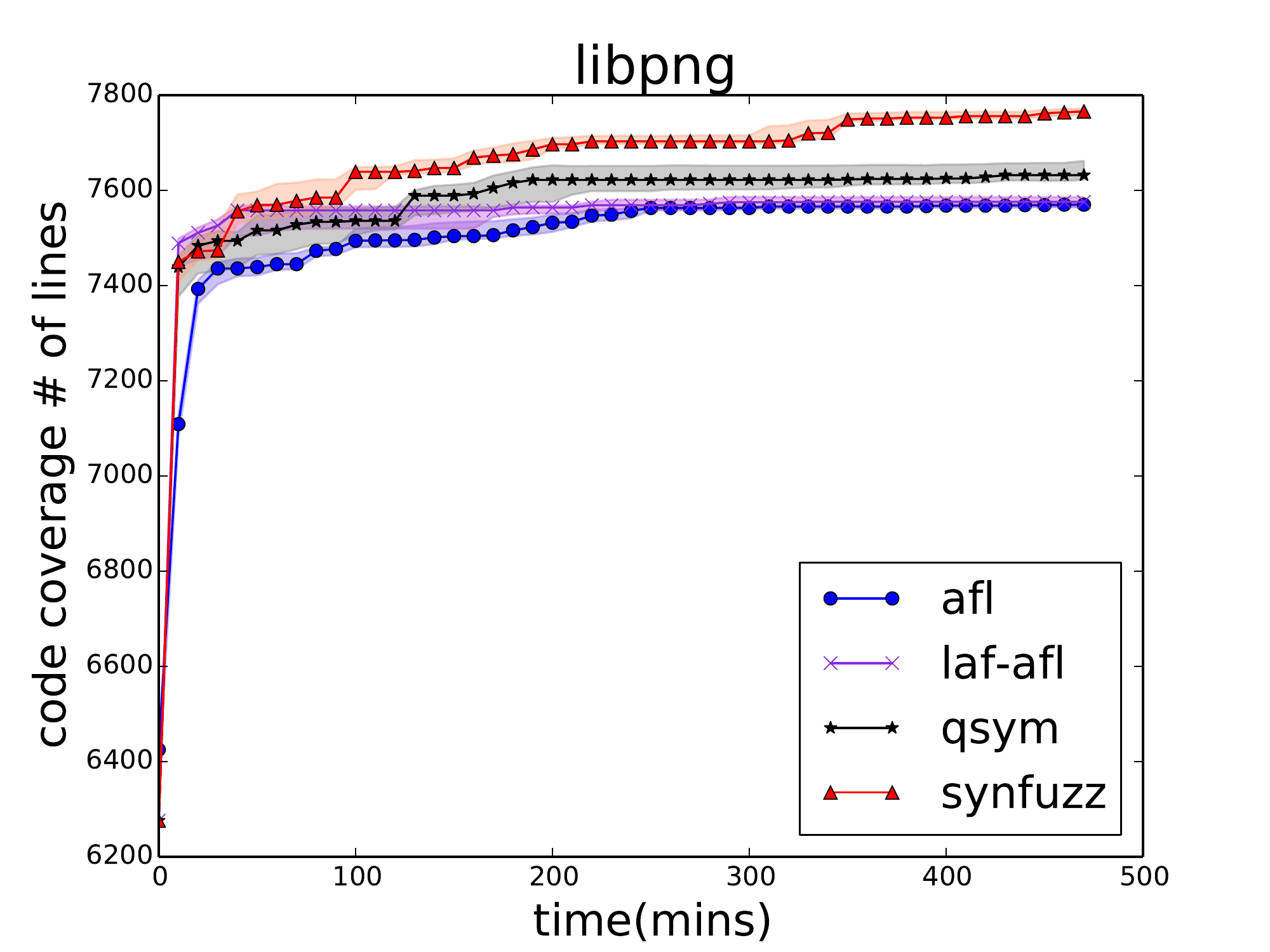}
  \end{subfigure}
  \begin{subfigure}[t]{0.23\textwidth}
    \includegraphics[width=\textwidth]{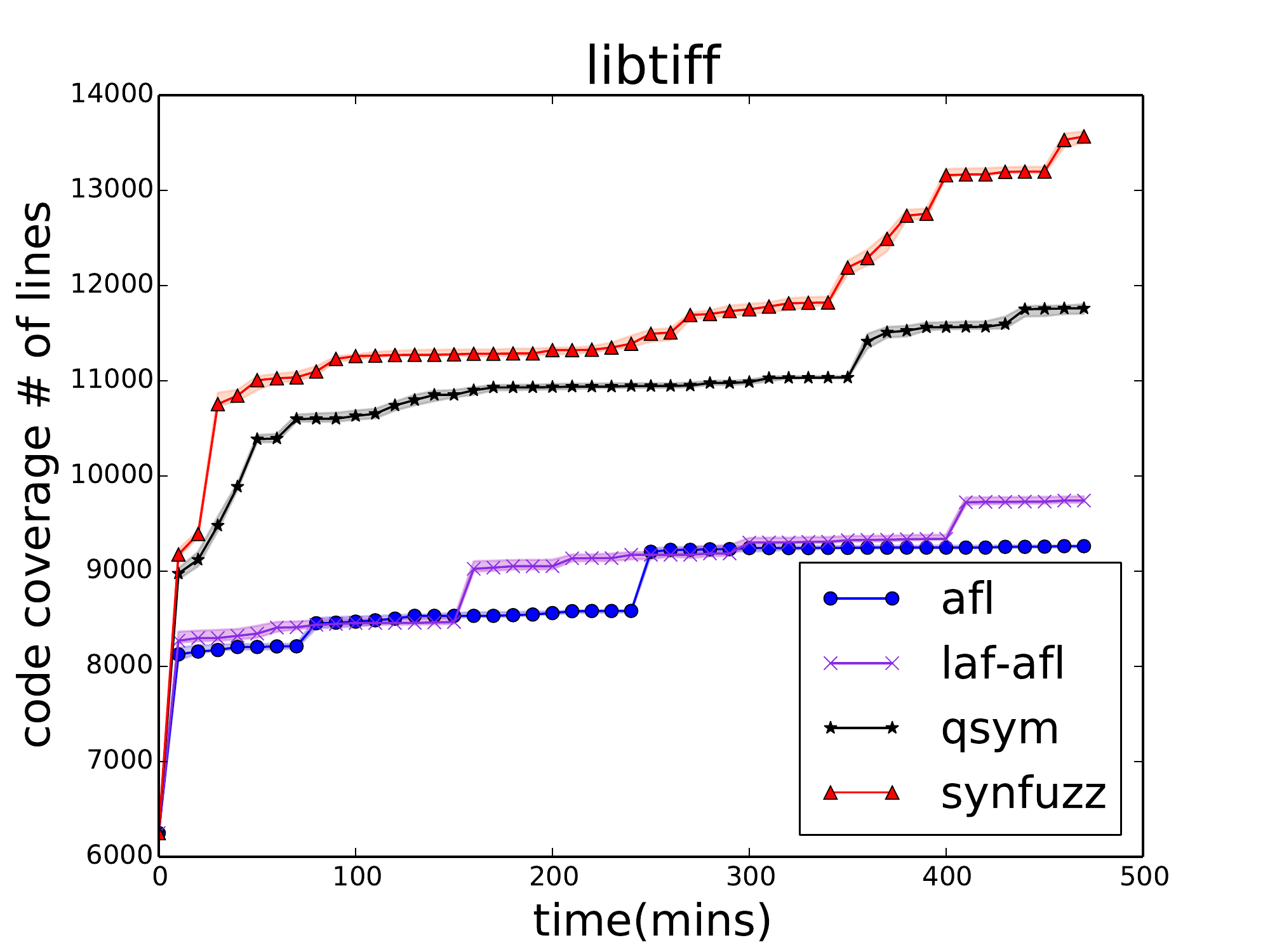}
  \end{subfigure}
  \begin{subfigure}[t]{0.23\textwidth}
    \includegraphics[width=\textwidth]{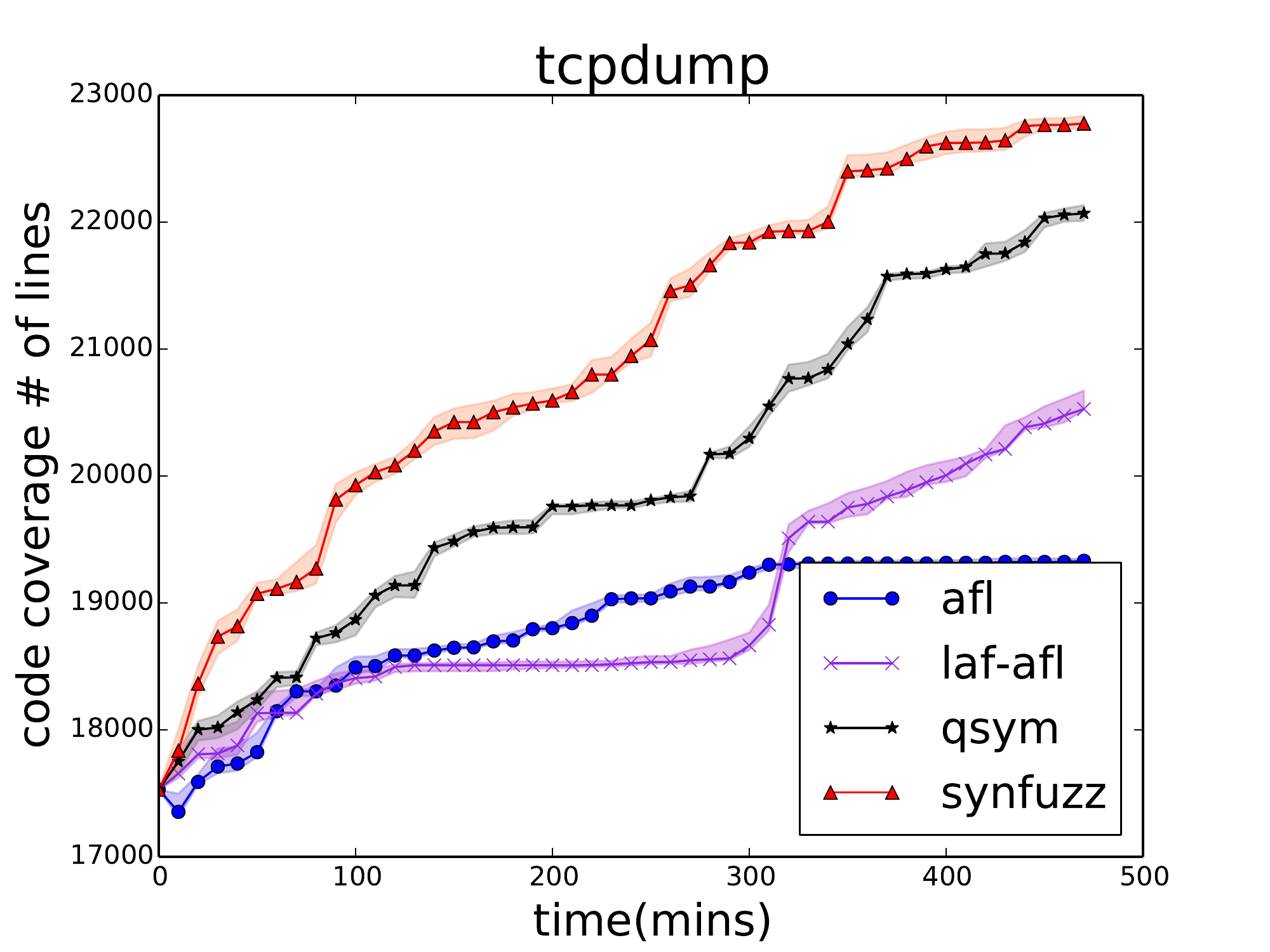}
  \end{subfigure}
  \caption{Code coverage comparison with different fuzzers. The x-axis
    shows the fuzzed time in minutes and y-axis shows the average line
    coverage. Shades around the lines are confidence intervals.}
  \label{fig:realworld}
  \vspace{-2em}
\end{figure}

\PP{Experimental Setup}
All our evaluations were performed on a server with an Intel(R) Xeon(R)
E5-2683 v4 @ 2.10GHz (40MB cache) and 512GB of RAM, running Ubuntu 16.04
with Linux 4.4.0 64-bit.

\subsection{Scalability}
\label{s:eval:scale}

To evaluate whether \sys can scale to real-world applications and
correctly synthesize symbolic formulas for branch predicates,
we used complex softwares which are tested by fuzzers for a long time.
We have picked some softwares tested by AFL and OSS-Fuzz: \cc{libjpeg},
\cc{libtiff}, \cc{openjpeg}, \cc{tcpdump}, \cc{file}, and \cc{binutils}
and libraries in FTS.
It is worth to note that most concolic execution engines cannot run
most of the applications that \sys was able to run~\cite{yun:qsym}.

In order to support these applications, we implemented several
features to \sys.
More specifically, because DFSAN performs source-code-based instrumentation,
we cannot perform a correct taint analysis for code inside an uninstrumented library.
The default way is to write a customize/wrapper function for the corresponding
library function.
However, since FTS benchmarks depend on many libraries, this manual approach
does not scale very well.
To solve this problem, we compiled all the required libraries with DFSAN.
For standard C library, we used customized/wrapper functions to propagate
taint labels. For standard C++ library, we used instrumented \cc{libc++}.

We have measured code coverage of \sys, and compared it with recent works:
AFLFast~\cite{bohme2016coverage} (baseline), LAF~\cite{laf-intel},
and QSYM~\cite{yun:qsym}.
\sys uses parallel fuzzing to sync new inputs from AFL, and tries
to mutate input bytes which affect target branch conditions.
With gathered inputs, condition values, and a union table, \sys generates
inputs that may flip the target branches, and test the inputs that actually
flip the target branches. If an input is able to flip the target branch,
\sys checks whether the input covers new edges by comparing AFL's hitmap.
If it explores a new edge, the input remains in its own queue and will be
synchronized to AFL fuzzers. This parallel fuzzing is similar to that of QSYM,
and we also use three processes: two for AFL fuzzer and one for \sys.
Since \sys and QSYM use three processes while others use one, we run
\sys and QSYM for 8 hours with 3 cores and others for 24 hours with 1 core.
We ran 10 times for each application.
If the benchmark contains seed inputs, we used the provided seeds;
otherwise we used an empty seed.
For coverage improvement, we followed the suggestion from~\cite{klees2018fuzzeval},
we performed Mann-Whitney U Test with standard 2-tailed hypothesis test.
When the $p$-value is less than 0.05, the improvement is significant.

\PP{Real-world applications.}
\autoref{table:realworld} shows the evaluation results on real-world
applications.
Overall, \sys is able to improve the coverage on all applications over
AFLFast, LAF, and QSYM.
\autoref{fig:realworld} shows the line coverage changes during
fuzzing. As shown, \sys also achieved higher code coverage faster.

\PP{FTS.}
\autoref{table:fts} shows the evaluation results.
Overall, \sys is able to improve the coverage on 15 out of the 26 benchmarks
when compared with AFLFast,
including all benchmarks marked as coverage testing:
\cc{freetype2} (22.82\%), \cc{libjpeg-turbo} (13.96\%), \cc{libpng} (10.36\%),
and \cc{wpantund} (3.93\%, when multi-branch solving is disabled).
Comparing with QSYM, \sys has better coverage on 11 benchmarks,
including all four benchmarks marked as coverage testing;
and has less coverage on 5 benchmarks.

For benchmark where we did not observe significant coverage improvement,
we performed additional analysis.
Some of these cases are due to the fact that the target tests are
relatively simple (\cc{c-ares}, \cc{json}, \cc{llvm-libcxxabi}, and \cc{re2})
so the line coverage is easily saturated.

\begin{figure*}[t]
  \centering
  \begin{subfigure}[t]{0.24\textwidth}
    \includegraphics[width=\textwidth]{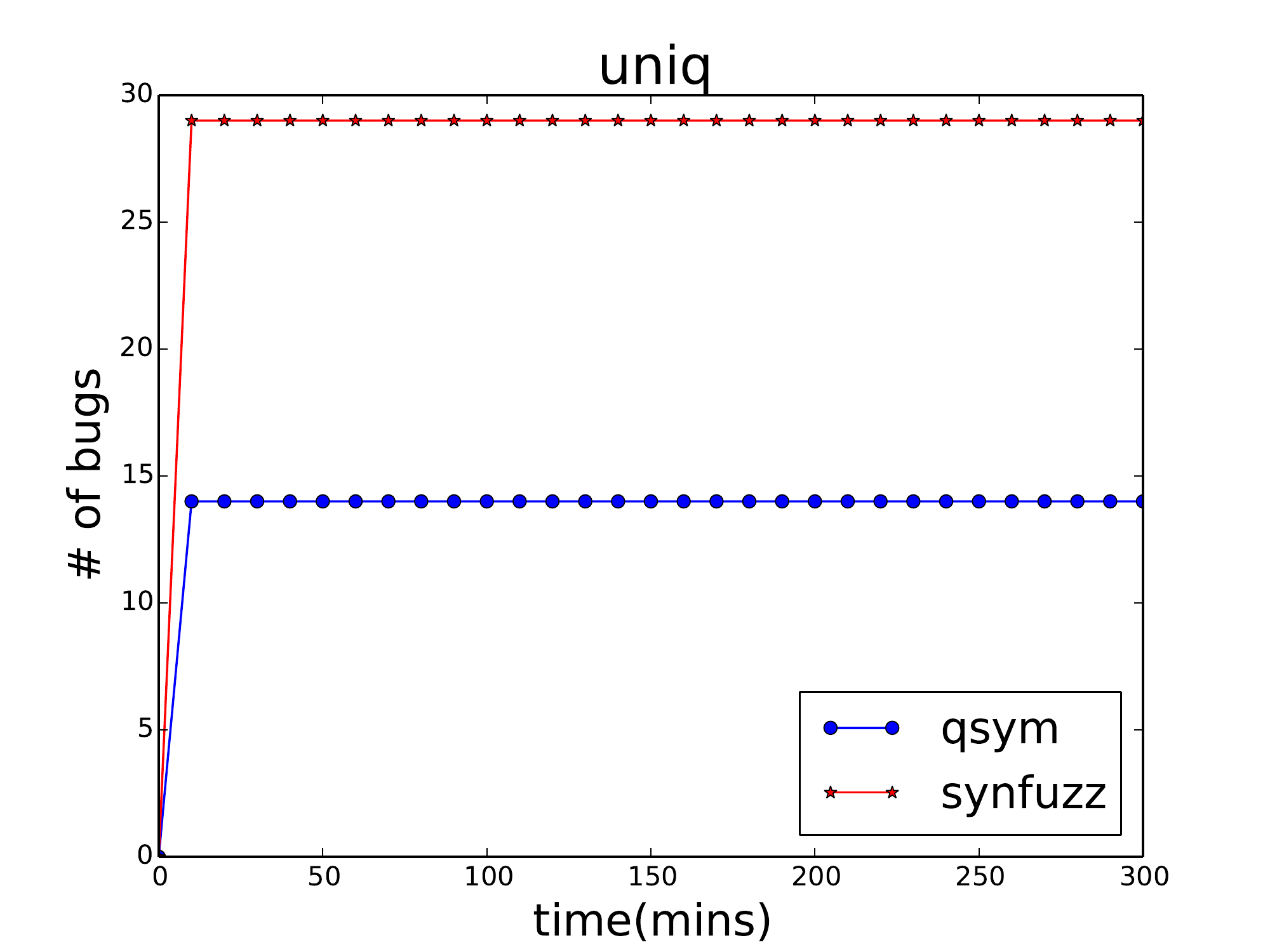}
  \end{subfigure}
  \begin{subfigure}[t]{0.24\textwidth}
    \includegraphics[width=\textwidth]{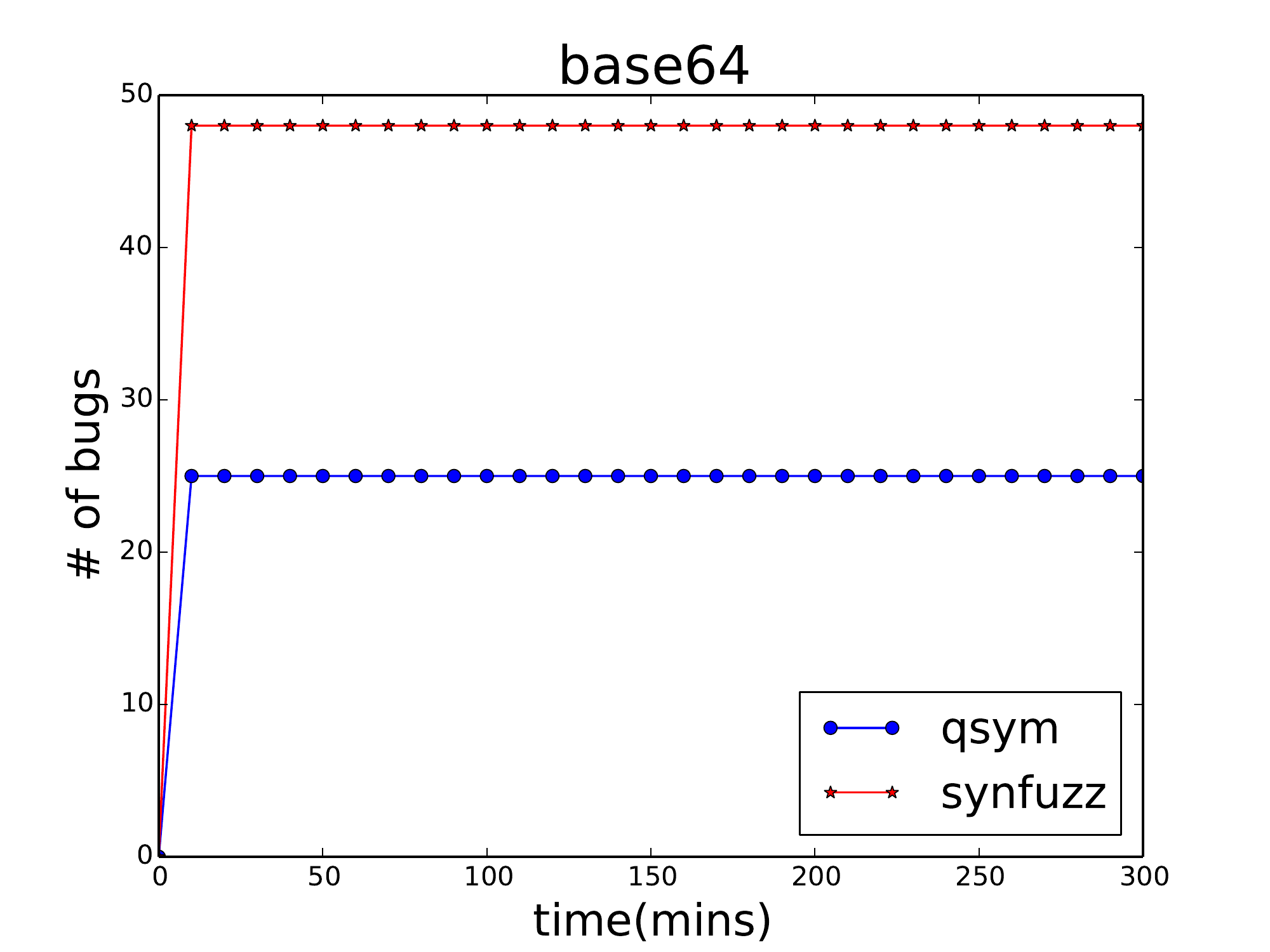}
  \end{subfigure}
  \begin{subfigure}[t]{0.24\textwidth}
    \includegraphics[width=\textwidth]{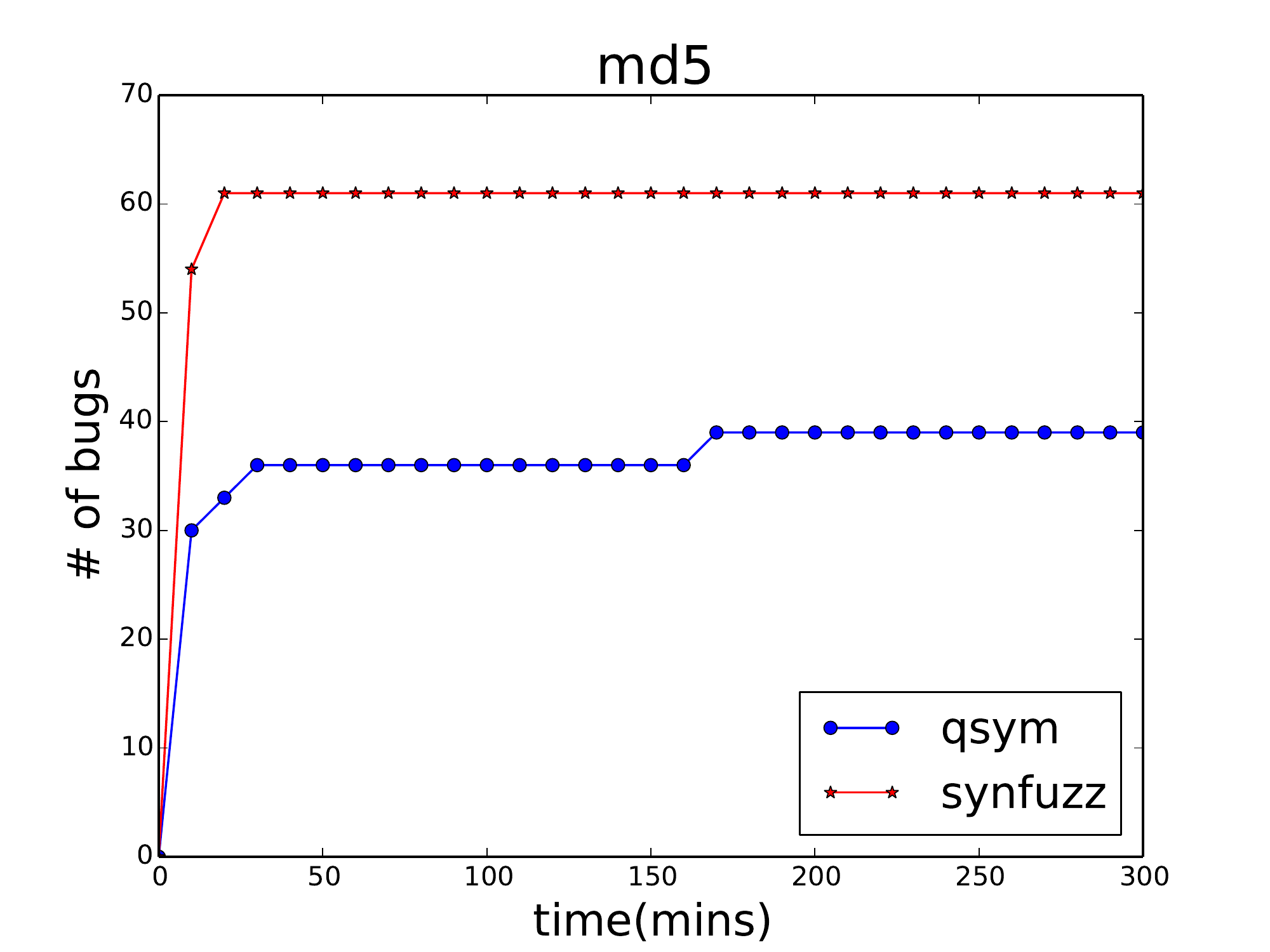}
  \end{subfigure}
  \begin{subfigure}[t]{0.24\textwidth}
    \includegraphics[width=\textwidth]{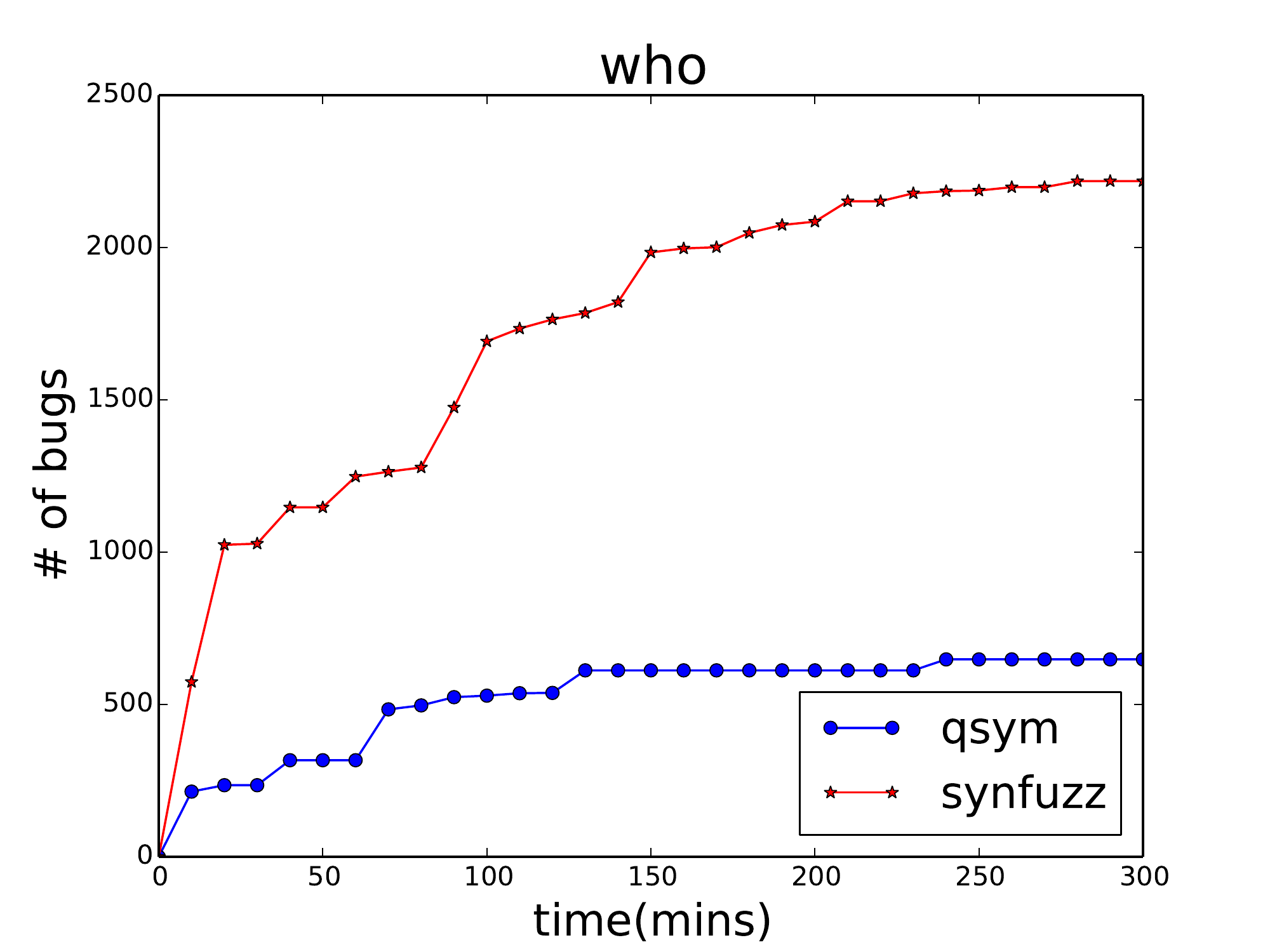}
  \end{subfigure}
  \vspace{-0.5em}
  \caption{Bug finding time of \sys compared to QSYM for LAVA-M benchmarks.}
  \label{fig:lava}
  \vspace{-2em}
\end{figure*}
\subsection{Bug finding.}
\label{s:eval:bugs}

To evaluate whether \sys effectively synthesizes branch constraints
and reduces time to find bugs, we used the LAVA-M and FTS benchmarks.

\PP{LAVA-M.}
LAVA is a technique to inject a large number of hard-to-find bugs into
program source code~\cite{dolan2016lava}.
The LAVA-M corpus is created by injecting multiple bugs into four
GNU coreutils programs.
Each injected bug has its own triggering condition and a unique ID
will be printed.
Each benchmark program is distributed with an initial seed and which
command line option should be used.
In this evaluation, \sys is configured to use one process,
fidgety mode, no dictionary, with fork server,
and with multi-branch solving.
Each benchmark was run for \emph{five} hours, which is the test
duration in the original LAVA paper~\cite{dolan2016lava}.

\autoref{tbl:lava-m} compares the bugs found by \sys with data
reported by other fuzzers and concolic execution engine
\cite{angora,yun:qsym,vuzzer,li2017steelix,peng2018t,
aschermann2019redqueen,you2019profuzz,she2018neuzz}
and by the original LAVA paper~\cite{dolan2016lava}.
Overall, \sys outperformed all state-of-art systems except
\textsc{Redqueen}~\cite{aschermann2019redqueen}.
For benchmarks with fewer bugs (\cc{uniq}, \cc{base64}, \cc{md5sum}),
\sys is on par with state-of-the-art systems---they all found
all the listed bugs as well as some additional ones.
The last benchmark \cc{who} has many more bugs than the other three
programs and is the most challenging one.
On this benchmark, \sys is second to \textsc{Redqueen},
who was able to find all the bugs and 328 more.
But \sys is able to solve more complex branch predicates.
In 5 hours, \sys was able to find 1,958 (91.67\%) listed bugs and
270 unlisted bugs.
It worth noting that standard LAVA-M benchmarks are compiled into 32-bit
binaries; however, because DFSAN does not support 32-bit mode,
we have to compile them into 64-bit modes.
We have manually changed the build script in LAVA, to be supported by \sys.
This may introduce some discrepancies, as some bugs cannot even be
validated using the inputs provided as part of the benchmarks.

\autoref{fig:lava} compares \sys with QSYM on the number of bugs found
by fuzzing time.
\sys is very fast on the simpler ones, it only takes less than 10 minutes
to reach all found bugs in \cc{uniq} and \cc{base64},
and less than 20 minutes for \cc{md5sum}.
On \cc{who}, QSYM finds significantly less bugs than the reported number
in~\cite{yun:qsym}, which is likely to be caused by 64-bit mode.

\begin{table}[t]
  \centering
  \caption{Reaching time of \sys and AFLFast to known bugs/code position.}
  \begin{tabular}{llrr}
\toprule
\multirow{2}{*}{App} & \multirow{2}{*}{Location} & \multicolumn{2}{c}{Reaching Time (hours)} \\ \cmidrule{3-4}
      &     &  AFLFast & SynFuzz \\
\midrule
boringssl-2016-02-12  & asn1_lib.c:459         & 2.97 & 3.21 \\
c-ares-CVE-2-16-5180  & area_..._query.c:196   & 0.02 & 0.09 \\
guetzli-2017-03-20    & output_image.cc:398    & 2.29 & 1.27 \\
json-2017-02-12       & fuzzer-..._json.cpp:50 & 0.04 & 0.01 \\
lcms-2017-03-21       & cmsintrp.c:642         & 3.14 & 2.54 \\
libarchive-2017-01-04 & archive_..._warc.c:537 & 5.78 & 4.52 \\
libpng-1.2.56         & png.c:1035             & 6.15 & 5.97 \\
pcre2-10.00           & pcre2_math.c:5968      & 0.35 & 0.26 \\
\bottomrule
\end{tabular}

  \label{table:bugfound}
  \vspace{-1em}
\end{table}

\PP{FTS.}
\autoref{table:bugfound} shows the time \sys and AFLFast took to
find bugs or reach a certain line of code in several benchmarks in FTS.
In this test, \sys did not perform as well as AFLFast in 2 of the 8 applications.
For these applications, the target bugs are behind a branch that can be easily
solved through mutation.
Therefore, \sys does not have an advantage in these particular cases.
However, in the applications where the bugs are behind complex and/or tight
branch conditions, \sys can finds target bugs faster than AFLFast.

\begin{figure}[ht]
  \centering
  \begin{subfigure}[t]{0.23\textwidth}
    \includegraphics[width=\textwidth]{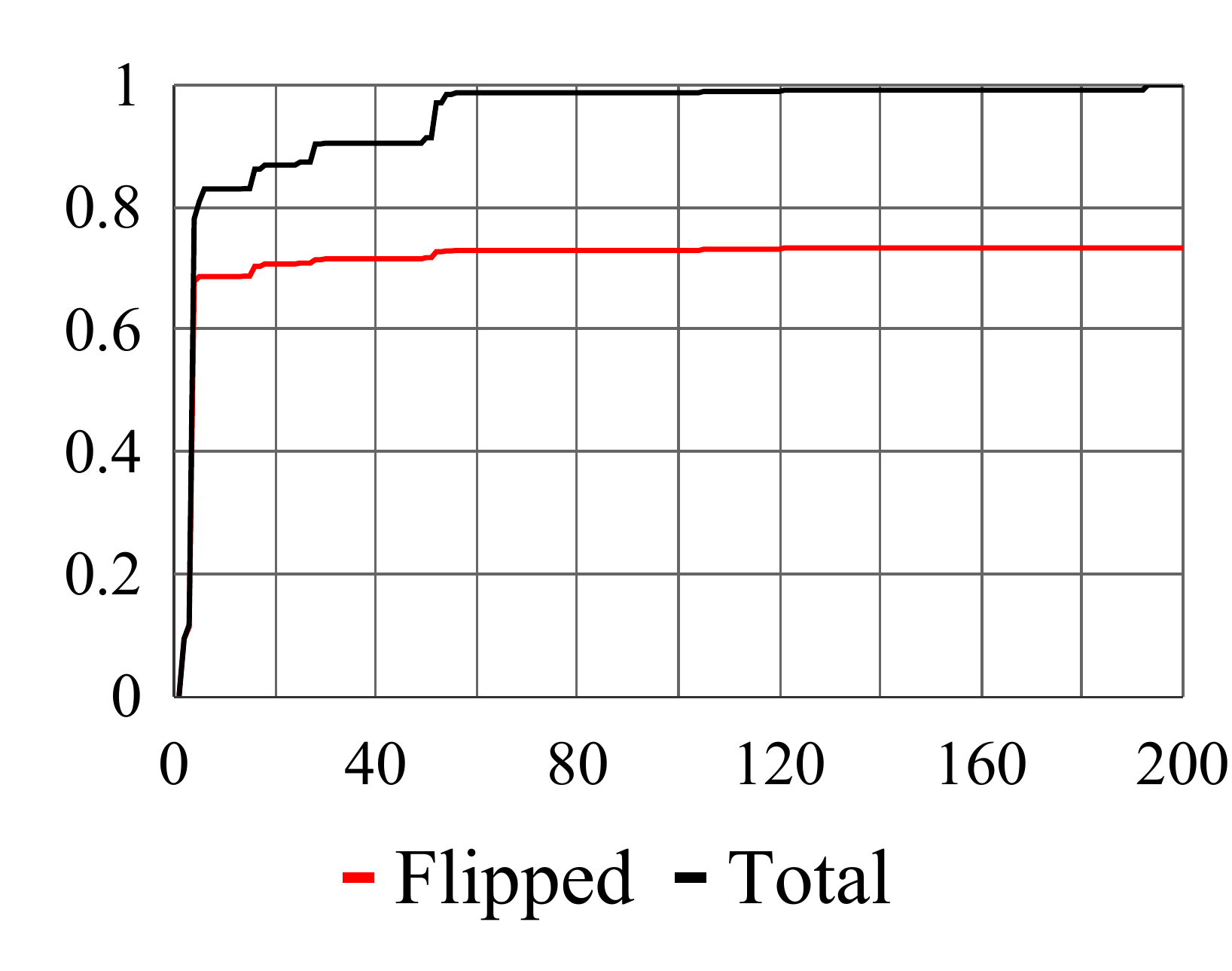}
    \caption{Lines}
    \label{fig:lines}
  \end{subfigure}
  \begin{subfigure}[t]{0.23\textwidth}
    \includegraphics[width=\textwidth]{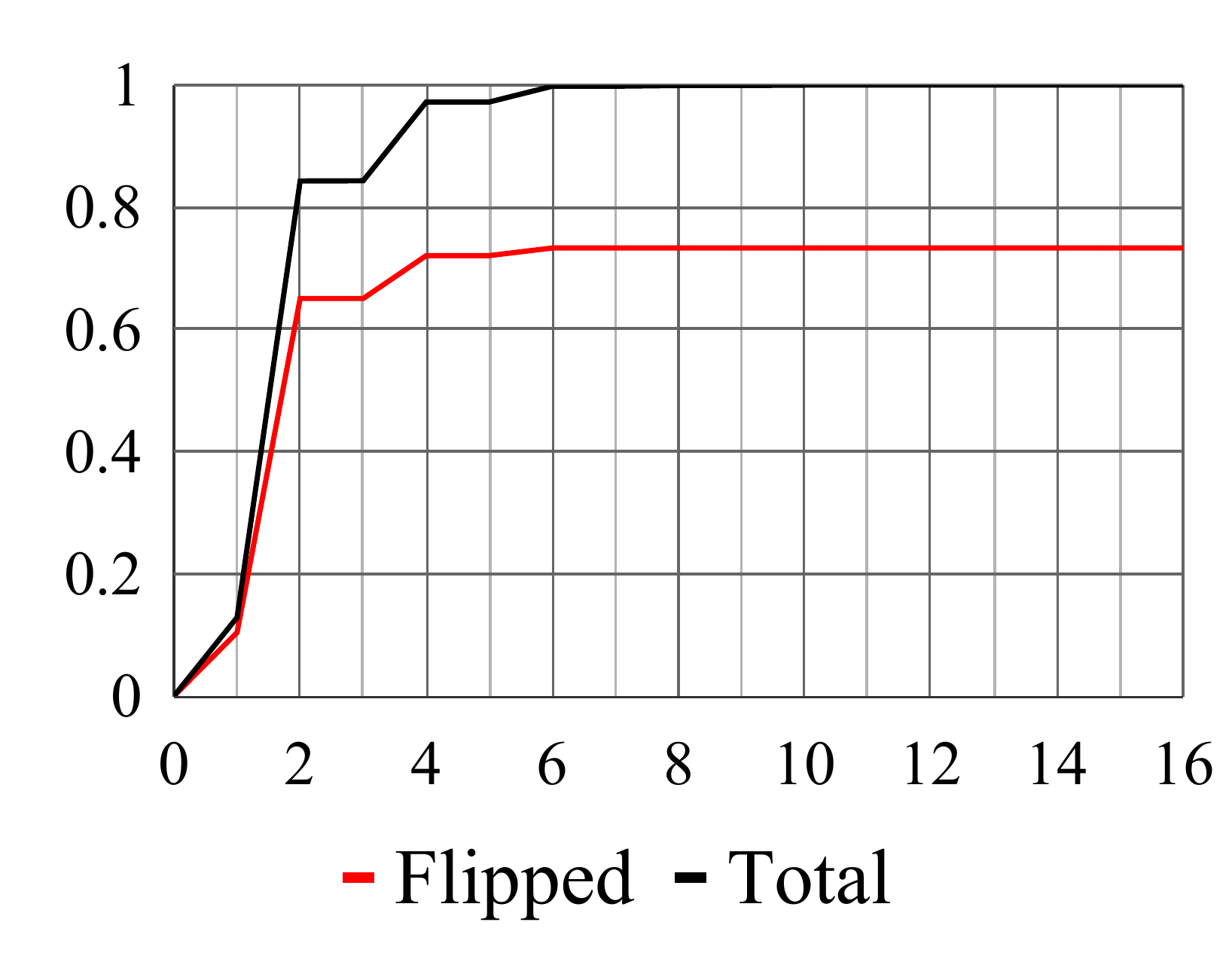}
    \caption{Involved input bytes}
    \label{fig:bytes}
  \end{subfigure}
  \vspace{-0.5em}
  \caption{Cumulative distribution of the number of lines and the number of involved input bytes
           of the synthesized functions}
  \label{fig:cdf}
  \vspace{-2em}
\end{figure}

\subsection{Synthesis}
\label{s:eval:synthesis}
To validate our synthesis-based approach is effective at flipping
symbolic predicates, we evaluated the complexity of synthesized functions,
which reflects the complexity of the corresponding branch predicates.
Since LAVA-M has tight but not complex branch conditions,
we used real-world applications
\cc{file} and \cc{libtiff} for this evaluation.
\autoref{fig:cdf} shows the cumulative distribution of the number of lines
and involved input bytes of synthesized functions (in black)\footnote{
For display, we capped the maximum number of lines at 200.}.
On average, synthesized functions have 10.19 lines of instructions,
and 2.23 involved input bytes.
We also measured the successful rate of the synthesizer.
A symbolic predicate is considered to be correct if the generated
input indeed flipped the corresponding branch.
In ~\autoref{fig:cdf}, the red lines represent the number of correctly
synthesized functions.
With optimistic solving, \sys can correctly synthesize 73\% of
total functions.

\begin{figure}[ht]
  \centering
  \begin{subfigure}[t]{0.23\textwidth}
    \includegraphics[width=\textwidth]{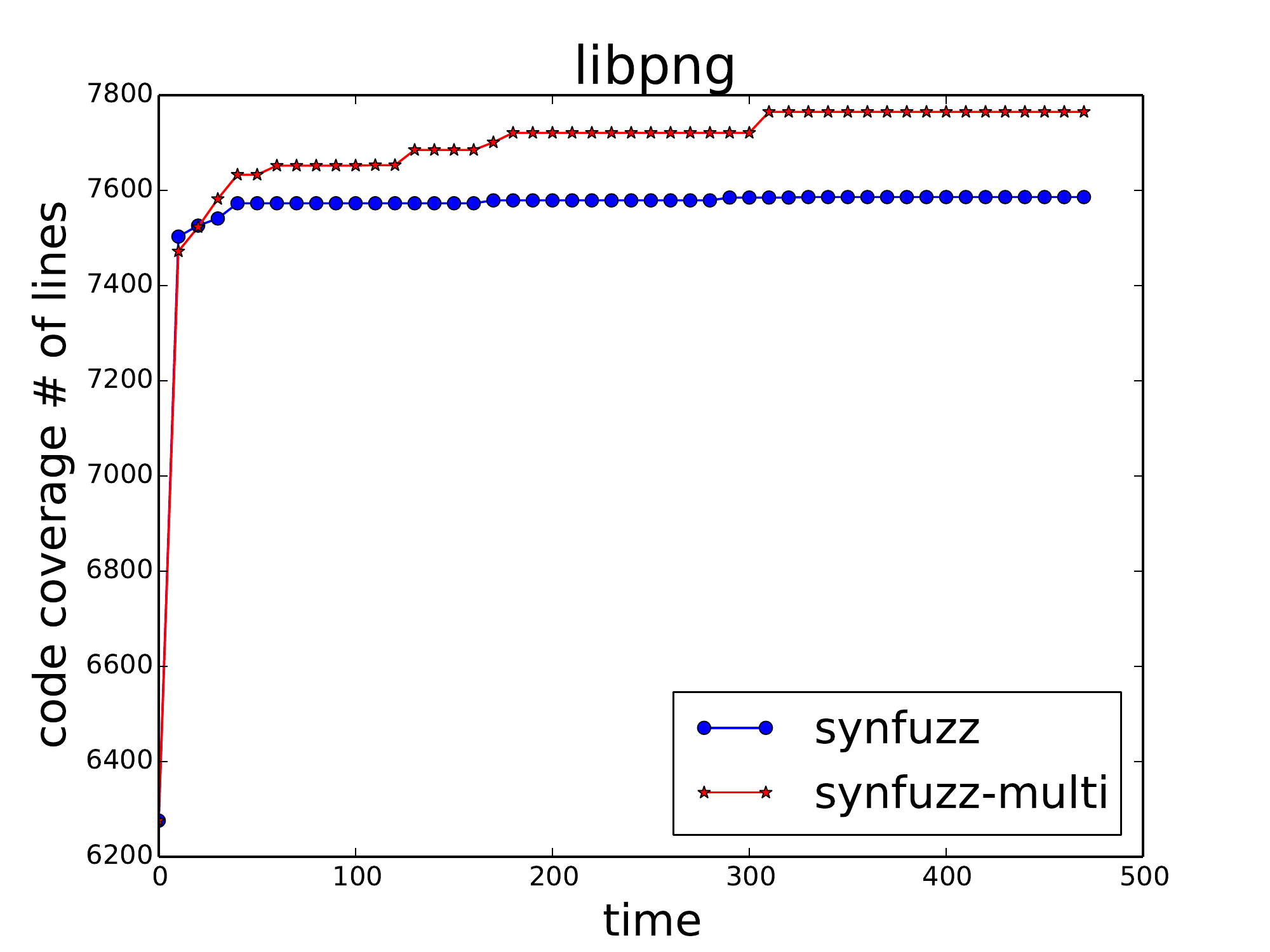}
    \caption{libpng}
    \label{fig:libpng}
  \end{subfigure}
  \begin{subfigure}[t]{0.23\textwidth}
    \includegraphics[width=\textwidth]{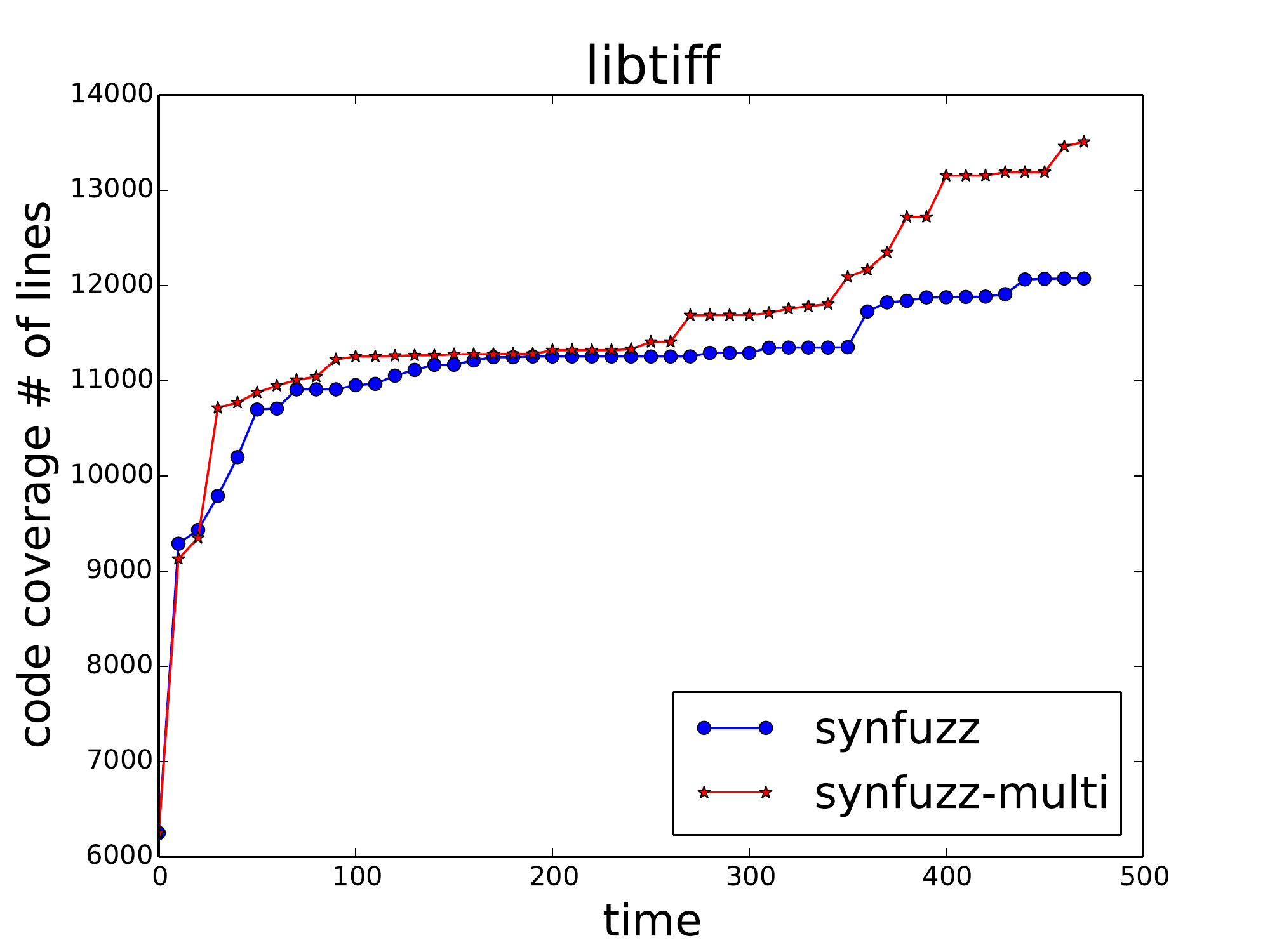}
    \caption{libtiff}
    \label{fig:libtiff}
  \end{subfigure}
  \vspace{-0.5em}
  \caption{Code coverage with and without multi-branch solving.}
  \label{fig:multi}
  \vspace{-2em}
\end{figure}

\subsection{Multi-Branch Solving}
\label{s:eval:nested}
Multi-branch solving was able to have benefit on flipping a nested branch,
which is especially beneficial for testing image processing applications,
since they have nested checks for image headers.
To see effectiveness of multi-branch solving, we used \cc{libpng} and
\cc{libtiff} \emph{without dictionary}.
\autoref{fig:multi} shows code coverage of \sys with and without
multi-branch solving.
\autoref{table:fts} also shows \sys's performance with and without
multi-branch solving.
Overall, although disabling multi-branch solving could make the concolic
execution faster and improve coverage on some applications,
enabling it still allows \sys to explore more code on most applications.

\section{Limitations and Future Work}
\label{s:discuss}


\PP{Supported Operations.}
First, \sys currently does not support floating point operations,
which could be added by tracking taint information for floating point
operations and leveraging solver's support for the floating point theory.
Second, \sys currently only considers branch operations
(\cc{icmp} and \cc{switch}) as taint sinks/synthesis targets.
For future work, we plan to support synthesizing operands of other
operations, such as pointer arithmetic~\cite{jia2017towards} and
memory allocation~\cite{wang2010taintscope}.

\PP{More Accurate Branch Tracking.}
Although we added context-sensitivity to branch tracking,
we still observed incorrect input-output pairs being collected due to
collision of branch id and context id and more importantly loops.
For example, consider a simple loop
\begin{center} \cc{while (i < 100) ++i;} \end{center}
In this case, the loop predicate can be expressed as $i + c$;
however, though $c$ is concrete, it is not a constant---after each iteration,
$c$ is increase by one.
However, because context-sensitive is not able to distinguish
different iterations, the synthesizer will not be able to generate
the symbolic formula; again, because $c$ is not a constant.
To solve this problem, we plan to explore path-sensitive branch tracking.

\PP{Binary-only Taint Tracking.}
Our current taint analyzer is based on DFSAN, which is included in the LLVM.
While this simplifies our implementation and is more efficient,
it also limits the capability of our prototype.
In particular, we cannot apply \sys to binaries or software that is
not compatible with DFSAN, such as the OS kernel.
To overcome this limitation, we could utilize other taint analyzers,
such as PIN~\cite{yun:qsym} and QEMU~\cite{henderson2017decaf}.

\PP{Faster Synthesis.}
Although we have already limited the search scope of synthesis to
concrete values, it still requires many inputs to find the correct
assignment when non-linear operations are involved.
One possible direction to further improve the speed of synthesis is to
use machine learning technique to learn from previous successful
results~\cite{balog2016deepcoder,lee2018accelerating,si2018learning}.

\section{Related work}
\label{s:relwk}

\PP{Concolic Execution}
Besides the performance issue, another challenge for concolic execution
is the path explosion problem.
To mitigate this problem, SAGE~\cite{godefroid2008automated} proposed
utilizing generational search to increase the number of generated
test cases in one execution, which has been adopted by most following
up work.
To compensate the scalability problem of concolic execution engines,
another popular approach is to combine concolic executing with fuzzing
~\cite{majumdar2007hybrid,stephens2016driller,yun:qsym,zhao2019send}.
In these approaches, path exploration is mostly done by the fuzzer,
who is more effective at exploring easy-to-flip branches.
Whenever the fuzzer encounters a hard-to-flip branch, it asks the
concolic execution engine to solve it.
The closest work to \sys is QSYM~\cite{yun:qsym}, which also focuses
on improving the scalability of concolic execution by minimizing
the use of symbolic interpretation.
Compare to QSYM, our approach is more ``extreme:'' we completely
eliminated the use of symbolic interpretation and resort to program
synthesis to construct the symbolic formula of branch predicates.
Neuro-Symbolic execution~\cite{shen2019neuro} explores another interesting
direction---learning a neuro-representation of branch predicate based on
input-output pairs and use optimization techniques to flip the branch.
Compare to \sys, this method can solve constraints that are not
solvable by a SMT solver; however, it also requires large amount of
training data and is slower.
So, we would argue it is complementary to \sys.

\PP{Taint-guided Fuzzing}
Dynamic taint analysis (DTA) another popular technique to improve the efficiency of fuzzing.
TaintScope~\cite{wang2010taintscope} utilizes DTA to discover and bypass checksum checks and
target input bytes that can affects security system library call.
Vuzzer~\cite{vuzzer} uses DTA to locate magic number checks then changes
the corresponding input bytes to match the magic number.
Steelix~\cite{li2017steelix} also uses DTA to bypass magic number checks but has better heuristics.
Redqueen~\cite{aschermann2019redqueen} uses the observation that input byte could indirectly
end up in the program state (memory), so by directly compare values used in compare instructions,
it is possible to infer such input-to-state relationships without expensive taint tracking.
Our evaluation results re-validated this observation as we found many 1-line symbolic predicates.
However, for complex predicates, this simple method will not work.
Neuzz~\cite{she2018neuzz} approximates taint analysis by learning the input-to-branch-coverage
mapping using neural network, which can then predict what inputs bytes can lead to more coverage.
However, it still uses mutation-based solving so will have problems solving complex branch predicates.
Eclipser~\cite{choi2019grey} exploits the observation that many branch predicate are either linear or
monotonic with regard to input bytes and solves them using binary search.
Angora~\cite{angora} is another close approach to \sys.
Its numeric gradient descent based solving is more general than flipping
simple magic number check.
Compare with Angora, \sys have two advantages:
(i) \sys considers nested branches when flipping a branch and
(ii) for simple predicate, synthesis requires less input-output pairs.

\PP{Learning-based Fuzzing}
In recent years, machine learning, especially deep learning has also been widely adopted in fuzzing.
Learning-based fuzzing approaches learn from large amount of valid inputs and use the learned model to guide the mutation,
so it is more likely to generate valid inputs.
This approach is much more efficient when the input has well structured format and
the target program employs rigorous input checks.
Skyfire~\cite{wang2017skyfire} uses data-driven seed generation technique
to generate the well-structured seed inputs from huge amount of previous samples.
Learn\&Fuzz~\cite{godefroid2017learn} uses sample inputs and neural-network-based
statistical machine learning techniques to generate of an input grammar.
GLADE~\cite{bastani2017synthesizing} uses synthesized grammar learned from provided seeds to generates inputs.
Compare with learning-base approaches, we believe \sys is more transferable,
\ie, it can be easily applied to a new program without losing its efficiency.

\section{Conclusion}
\label{s:conclusion}

In this paper, we proposed a new approach to perform concolic execution.
Our approach first uses dynamic taint analysis to capture a partial AST
of a branch predicate then uses oracle-guided program synthesis
to recover the full symbolic formula.
By doing so, we can eliminate the symbolic interpretation of instructions
thus greatly improved the scalability of concolic testing.
Our evaluation showed that our prototype \sys outperformed
many state-of-the-art testing techniques.

\balance
\bibliographystyle{IEEEtranS}
\setlength{\bibsep}{3pt}
\bibliography{conf,p}

\end{document}